\documentclass[aps,superscriptaddress,preprintnumbers,twocolumn,prl,showpacs]{revtex4-1}
\usepackage{graphicx}% Include figure fiiles
\usepackage{dcolumn}% Align table columns on decimal point
\usepackage{bm}% bold math
\usepackage{epsfig}
\usepackage{color}

\usepackage{longtable}
\usepackage{amsmath}
\usepackage{multirow}

\begin{document}

%\preprint{Submitted to {\it Physical Review B}}

%\title{Dislocation-mediated premelting of single-crystal copper under shock compression}
\title{Plasticity and the corresponding mechanisms of nanopowdered Mg during shock consolidation and spallation}

\author{D. B. He}
\affiliation{College of Science, Hunan Agricultural University, Changsha 410128, People's Republic of China}
%\affiliation{\mbox{Key Laboratory of Advanced Technologies of Materials, Ministry of Education,} Southwest Jiaotong University, Chengdu, Sichuan, 610031, People's Republic of China}

\author{M. Y. Wang}
\affiliation{College of Science, Hunan Agricultural University, Changsha 410128, People's Republic of China}

\author{W. B. Bi}
\affiliation{Laboratoroy of Computational Physics, Institute of Applied Physics and Computational Mathematics, PO Box 8009, Beijing 100088, People's Republic of China}

\author{M. Shang}
\affiliation{College of Science, Hunan Agricultural University, Changsha 410128, People's Republic of China}

\author{Y. Cai}
\affiliation{The Peac Institute of Multiscale Sciences, Chengdu, Sichuan, 610207, People's Republic of China}

\author{L. Deng}
\affiliation{College of Science, Hunan Agricultural University, Changsha 410128, People's Republic of China}

\author{X. M. Zhang}
\affiliation{College of Science, Hunan Agricultural University, Changsha 410128, People's Republic of China}

\author{J. F.~Tang}
\affiliation{College of Science, Hunan Agricultural University, Changsha 410128, People's Republic of China}

\author{L. ~Wang}
\email{Corresponding author. wangliang0329@hunau.edu.cn}
\affiliation{College of Science, Hunan Agricultural University, Changsha 410128, People's Republic of China}

\date{\today}

\begin{abstract}
Nanopowder consolidation under high strain rate shock compression is a potential method for synthesizing and processing bulk nanomaterials. A thorough investigation of the shock deformation of powder materials is of great engineering signicance. Deformation twinning plays a vital role in accommodating plastic deformation of np-Mg which is hexagonal close packed (hcp) metals, but its mechanisms are still unsettled under high strain rate shock compression. Here we combine nonequilibrium molecular dynamics (NEMD) simulations and X-ray diffraction (XRD) simulation methods to investigate the deformation twinning and pore compaction in shock-compressed np-Mg. Significant anisotropy and strong dependence on crystallographic orientation are presented during shock-induced deformation twinning. During the shock stage, three typical types of twins were firstly induced, namely $\{11\bar{2}1\}$ twin (T1), $\{11\bar{2}2\}$ twin (T2) and $\{10\bar{1}2\}$ twin (T3). Most of them were generated in grains with a larger angle between the impact direction and the c-axis of the lattice. With the increase of strain rate, the types and quantities of twins continued to enrich, but they did not occur when the strain rate ($u_{\rm p}$ $\ge$ 2.0 km\,s$^{-1}$) was too high. We also discussed the deformation mechanisms of the three types of twins and found that the coupling of slip and shuffle dominated twin deformation. In addition, void filling occurred due to the interaction of twinning and other plastic deformations, leading to the densification of np-Mg. During the release stage, an interesting reverse change was observed, where the twins produced by the impact receded, and twins were produced in grains that were previously difficult to produce.
\end{abstract}

\vspace{\baselineskip}
%\end{frontmatter}
\maketitle

\section{Introduction}
Ultrafine grained (UFG) solids with grain size ($D$) of 1 $\mu$m, and nanocrystalline (NC) materials with $D$ $<$ 100 nm \cite{chen03sci, hirth92b}, represent two distinct material types, owing to their unique physical properties, i.e., higher strength/hardness and toughness \cite{wang02apl, wang06am1, mo07apl}, excellent diffusivity \cite{baro01rams, kim00am, kolobov01sm}, and enhanced thermal expansion coefficients \cite{zhang98jap}. Such excellent properties and potentials prompt us to fabricate bulk UFG/NC materials, and two primary methods are developed, i.e., the top--down and bottom--up approaches \cite{ahn15msea}. For top--down approaches, the UFG/NC structures are produced via deforming grain refinement technique, i.e., the equal-channel angular pressing (ECAP) \cite{zhao16prl, langdon07msea, quang07jmpt} and high-pressure rolling (HPT) \cite{horita05msea, edalati16msea, azzeddine22pms}. For bottom--up approaches, UFG/NC materials are manufactured via powder metallurgy, such as inert gas condensation \cite{gleiter89pms, zheng20prm, patelli22jac}, spark plasma sintering \cite{hu20md, olevsky07jap, cavaliere19b}, and high pressure cold sintering \cite{guo19armr}. 

Shock consolidation \cite{schwarz84am, zhang08msea, gourdin86pms}, is a unique bottom--up method, by compacting nanopowders (NP) to strong bulk NC materials using shock waves from an explosive or high-speed collision \cite{jin04am, chen04jap, zhang22jap}. When shock waves pass through, it contributes to an ultra-high pressure in the powders. Then the particle undergoes an apparent particle deformation and densification, due to a collapse of powder agglomerates \cite{zhou16jac}. Compared to conventional methods, shock consolidation is one of the most efficient methods to manufacture bulk materials with improved properties \cite{meyers94b, thadhani93pms, viswanathan06mser}. For instance, the structure of the starting powders retains and no grain growth occurs during shock consolidation cite{meyers99am}, due to extraordinarily fast process within a microsecond ($\mu$s) time scale. Consequently, lots of bulk metallic/alloy nanostructured materials, such as aluminum \cite{nieh96am, venz03jms, brochu07msea, fredenburg10msea}, copper \cite{huang12jap}, iron \cite{dai08jap}, silver \cite{zhang08msea}, tungsten--copper \cite{zhou16jac, wang10msea}, high-entropy alloys \cite{yim17msea}, and aluminum ceramic \cite{bukaemskii08ces}, are fabricated using shock consolidations. However, the details of consolidation process, are rarely unveiled until now, which determine the structures and properties of final bulk nanostructured materials \cite{feng20jap}. In facts, capturing the microstructural features in dynamics and the corresponding mechanisms, is a key to optimize and control the dynamic powder compaction process.

Recently, both experiment and simulation efforts are devoted to understanding the mechanisms of shock consolidation. Experimentally, Matsumoto \cite{matsumoto89jms} found that molten jet, dynamic friction of particles, and the plastic deformation around a void, facilitate the compaction, using an analysis of scanning electronic microscope (SEM). It is validated by the following experiments on fabrication of fine-grained W-Cu composites \cite{zhou16jac}. In Kondo's experiment \cite{kondo90jacs}, it presented that local heating, induced by plasticity, accelerates the consolidation. Gao \cite{gao91jap} proposed a melting-welding consolidation mechanism based on the experiment for metal powder with a random arrangement. However, the technique difficulties for current set-ups, in probing the structural deformation on the atomic scale and ultrashort time scale \cite{liu19am}, limit the knowledge on the dynamic process during shock impact in depth. Recent developed femtosecond x-ray diffraction (XRD) using x-ray free-electron laser (XFEL) \cite{wenk13prl}, provides opportunities for determining the irreversible transient processes of plasticity deformation. However, we are not aware of such experiments in shock consolidation, although they have potentials for real-time, $in$ $situ$, and high-resolution temporal and spatial probes \cite{chen19prl, chai20ijp, turneaure20prl}.

Theoretically, the continuum models are implemented, to describe the compaction of powders in dynamics. They can help in both understanding the experiments and accessing the regimes that inaccessible in current experiments \cite{nieva14am}. In Boltachev's model \cite{boltachev09am} for granular medium, the evolution of boundaries between powder and the container, are taken into account in compaction model to describe the the mechanical response quantitatively. Ahn $et$ $al$. \cite{ahn15msea} then developed a Cu-compacted model by incorporating a dislocation model. They also found a strong dependence of dislocation for the propagation of shock wave and the void-collapse induced plastic flow. These models are appropriated to describ the deformation and its underlying mechanisms at mesoscale, but the validity of models in microscopic is not clear. For understanding plasticity and defect procedures of nanopowders during consolidation, at the microscopic level, MD simulations are employed. Huang $et$ $al$. \cite{huang12jap} studied shock consolidation of Cu nanopowders and the dynamic tensile response in their simulations. It is found that dislocation, lattice rotation, shearing, and friction contribute to the void collapse. Mayer et al. \cite{mayer20ijp} reported a strong relation between dislocation slips and compaction in aluminum nanopowder, and proposed a mechanical model for describing metal nanopowder compaction. Feng $et$ $al$. \cite{feng20jap} investigated the shock compaction of tungsten nanoparticles, showing a liquid-diffusion/solid-pressure welding mechanism during consolidation. 

Based on the results above, the shock-induced plasticity (i.e., dislocation slips) play the key role in shock consolidation, and thus an explicit description of deformation is necessary. However, the investigation of microstructure dependence of plastic deformation, i.e., its nucleation, dynamic evolution, and the corresponding mechanisms during shock compaction, is rare. Consequently, the goal of this work is to explore shock-induced plasticity in nanopowders, consisting of series of crystallographic orientations, and reveal the theoretical relationship between plasticity and microstructures.

\begin{figure}[t]
\centering
  \includegraphics[scale=0.35]{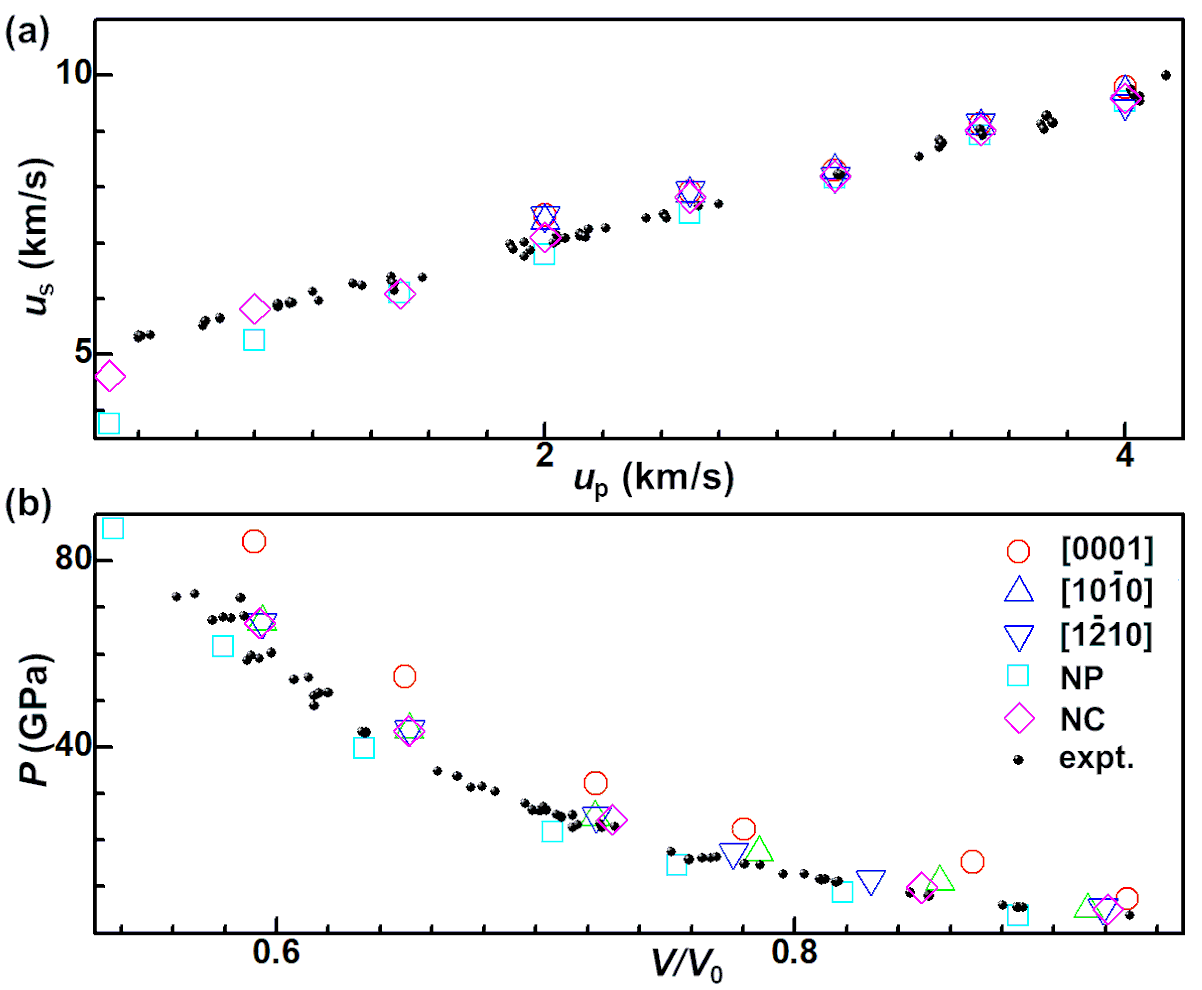}
  \caption{(a) $u_{\rm s}$ -- $u_{\rm p}$ plots obtained from our MD simulations (open symbols), and experiments (black dots), and (b) the corresponding plots of pressure vs normalized specific volume. Here NP and NC represent the nanopowdered and bulk nanocrystalline magnesium, respectively.}
  \label{eos}
\end{figure}

In this work, we systematically investigate the shock response and its microstructure deformation of magnesium (Mg) nanopowders during shock consolidation, by conducting nonequilibrium molecular dynamics (NEMD) simulations \cite{wang19prb} and $in$ $situ$ XRD analysis. Here, Mg has been chosen mainly because of its wide applications in aerospace, automotive, medical equipment, national defense science and industry \cite{chen15nat, lentz15nc, liu19sci, wu18sci}. This paper is organized as follows: a description of the details of MD simulations and related analysis methodology is presented in Sec.II, the results of simulations and discussion in Sec. III, followed by summary and conclusions in Sec. IV.

\section{Methodology}
For our NEMD simulations, the open source code Large scale Atomic/Molecular Massively Parallel Simulator (LAMMPS) \cite{plimpton95jcp}, and modified embedded atom method (MEAM) potential of Mg \cite{kim09cp}, are utilized. This potential can accurately describe the plasticity of Mg \cite{yi16am, tang23jma}, verified by density function theory \cite{ghazisaeidi14sm, wang22jma} and experimental results \cite{sun18jmps}. This potential also presents a reasonable accuracy for shock simulations. To check the performance of such MEAM potential during shock loading, we carry out simulations under different shock strengths in single- and nano-crystal Mg. For describing the effect of crystallographic orientation in single crystal, we choose the $x$-axis, parallel to the [1$\bar{2}$10], [10$\bar{1}$0], or [0001] directions as the impact direction. The dimensions of the single-crystal Mg are about 145 $\times$ 13 $\times$ 13 nm$^3$ ($\sim$ 1.2 million atoms), and the impact velocities $u_{\rm p}$ range from 1.0 to 4.0 km\,s$^{-1}$. Based on the $u_{\rm s}$--$u_{\rm p}$ relations of the plastic shocks (Fig.~\ref{eos}(a)), all simulation results, including single-crystals (shocked along [1$\bar{2}$10], [10$\bar{1}$0], and [0001], respectively), nanocrystals, and nanopowders, are in agreement with the experimental results \cite{baytos80b}. The functions of pressure vs normalized specific volume ($V$/$V_0$) are shown in Fig.~\ref{eos}(b). Despite the scatter in the experimental data, our simulation results are consistent with the experiments for the compression (1-$V$/$V_0$), ranging from 6$\%$ to 44$\%$. It presents a reasonable accuracy of this potential for shock simulations from the results above, although its accuracy for deformation at high pressure ($u_{\rm p}$ $>$ 4.0 km\,s$^{-1}$) remains to be established.

\begin{figure}[t]
\centering
\includegraphics[scale=0.3]{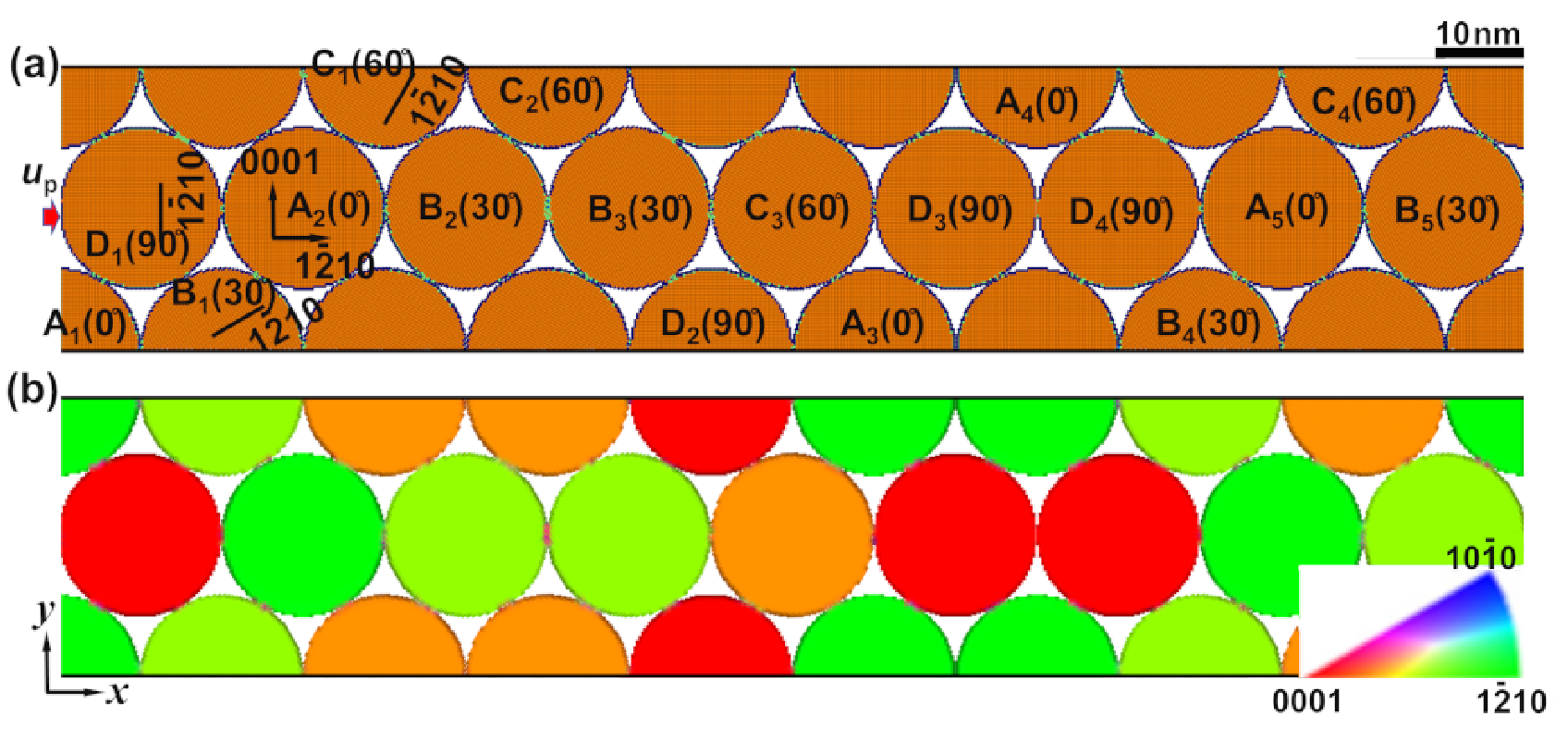} \label{model}
  \caption{ Atomic configurations of columnar NP-Mg projected onto (10$\bar{1}$0), colored with (a) CNA and (b) OM, respectively. The dimensions of the configuration are about 156 $\times$ 30 $\times$ 30 nm$^3$, consisting of about 5.6 million atoms. Shock direction: left $\to$ right, along the $x$ axis. A-, B-, C-, and D-type grains represent those oriented relative to the [1$\bar{2}$10] direction by an angle $\phi$ = 0$^{\circ}$, 30$^{\circ}$, 60$^{\circ}$, and 90$^{\circ}$, around $z$-axis ([10$\bar{1}$0] direction), respectively. Here, $u_{\rm p}$ is the piston velocity.}
\end{figure}

We construct an idealized columnar nanopowdered configuration, containing 18 grains of identical shape and diameter ($\sim$ 18 nm) with four different crystallographic orientations (A--D, Fig.~\ref{model}), in the three-dimensional (3D) periodic cell. The columnar axis is along [10$\bar{1}$0] crystallographic direction (the $z$-axis), and the thickness along this axis is about 30 nm. The grains are oriented relative to the $x$-axis ([1$\bar{2}$10] in grain A$_1$), by an angle of $\phi$ = 0$^{\circ}$, 30$^{\circ}$, 60$^{\circ}$, and 90$^{\circ}$ for grains A--D, respectively. The dimensions of the configurations are about 156 $\times$ 30 $\times$ 30 nm$^3$, and containing approximately 5.6 million atoms. The configurations are first relaxed at 0 K with the conjugate gradient method, and then thermalized at ambient conditions with a constant-pressure-temperature ensemble and 3D periodic boundary conditions, prior to shock loading. Shock simulations are then performed with the microcanonical ensemble. Periodic boundary conditions are applied along the $y$- and $z$-axes, while a free boundary is applied along the $x$-axis. The timestep for integration of the equation of motion is 1 fs. The small dense region on the left is set as the piston \cite{holian98sci} in our shock simulations. The interactions between the piston and the reset of the atoms in the configuration are described with the same interatomic potential, while the atoms in the piston do not participate in molecular dynamics. The atomic piston delivers the shock with the piston velocity of $u_{\rm p}$, ranging from 0.5 to 3.0 km\,s$^{-1}$, starting at time $t$ = 0 from $x$ = 0 along the $x$-axis towards the free surface. At a certain time ($t_{\rm c}$), the piston is stopped, creating a release fan propagating towards the free surface. The release fan then tend to interact with each other, due to shock reflection at the free surface, and gives rise to release and subsequent tension within the target interior \cite{luo09jap}.

The one-dimensional (1D) and two-dimensional (2D) binning analyses \cite{huang12jap}, resolving spatially physical properties such as stress $\sigma_{ij}$ ($i$, $j$ = $x$, $y$, and $z$), are performed to describe the response during shock loading. The bin width is 0.5 nm. The center-of-mass velocity $\bar{v}_i$ of a bin is removed when calculating $\sigma_{ij}$ within each bin: $\Delta\sigma_{ij}$ = -($m$/$V_a$)$\bar{v}_i\bar{v}_j$, where $m$ is the atomic mass and $V_a$ is the atomic volume averaged over the bin. Then the pressure $P$ and shear stress $\tau$ can be calculated \cite{wang19prb}. To characterize the structural deformation, the common neighbor analysis (CNA) \cite{tsuzuki07cpc, ackland06prb} and slip vector analysis \cite{wang14jap}, are implemented. To better visualize the plastic deformation in dynamics, the orientation mapping (OM) analysis, following the standard electron backscatter diffraction (EBSD) analysis, is implemented as a complement to atomic-level characterization. The simulated the X-ray diffraction (XRD) patterns \cite{wang15jap, bi23jap} are also calculated, to revel the the microstructure features and deformation in real time, by performing a parallel code of GAPD \cite{e18jsr}. Such XRD analysis is widely applied to quantitatively determine the microstructure characteristics \cite{chen19prl, zhang21prl, mo22nc, zhang23nc}, by using free electron lasers and synchrotron radiation in experimients.

\begin{figure*}[t]
\centering
  \includegraphics[scale=0.6]{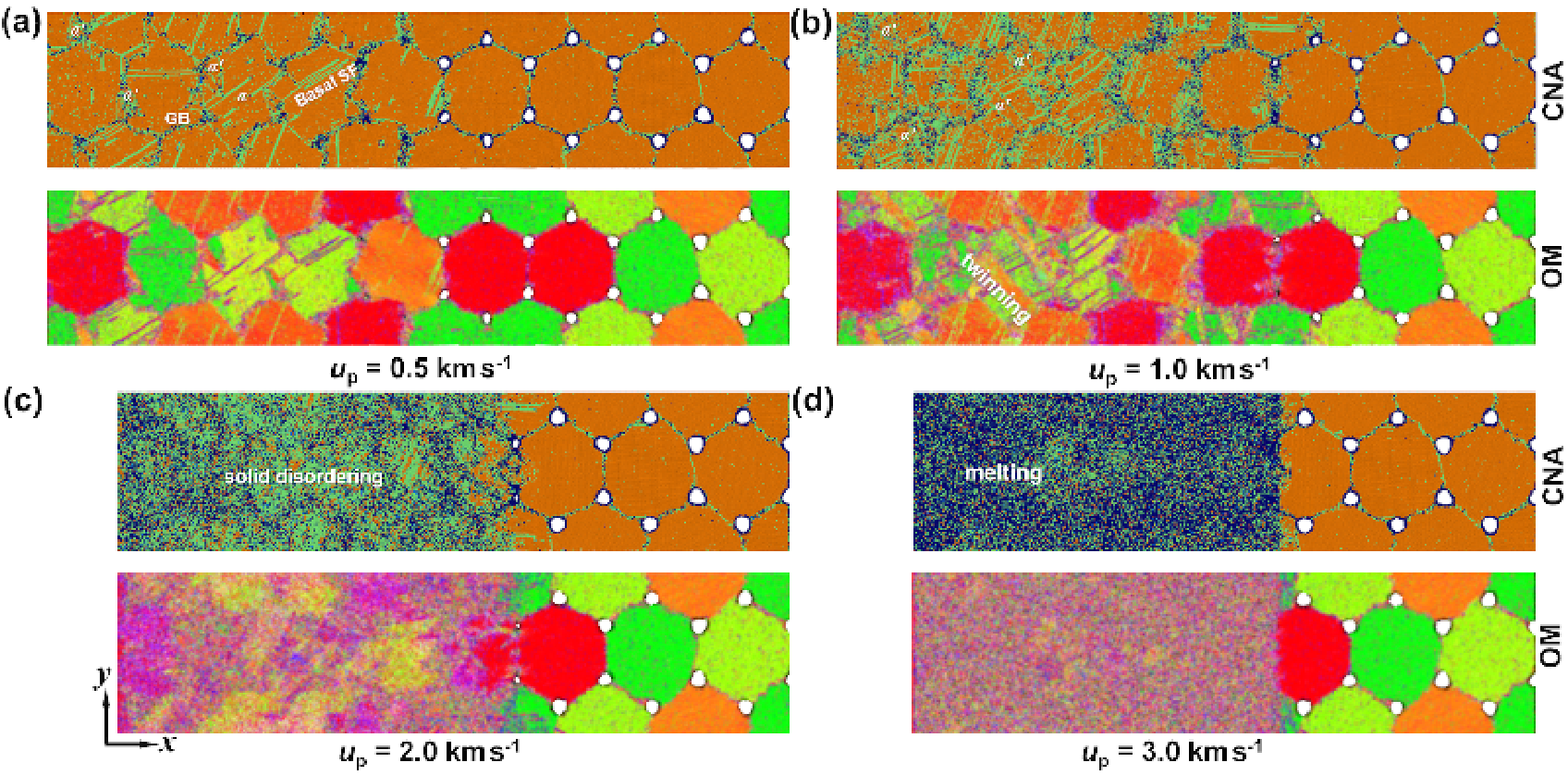}
  \caption{Atomic configurations, color-coded with common neighboring analysis (CNA) and corresponding orientation maps (OM), respectively, for the NP-Mg during shock compression, at (a) $u_{\rm p}$ = 0.5 km\,s$^{-1}$ ($t$ = 23 ps), (b) 1.0 km\,s$^{-1}$ ($t$ = 20 ps), (c) 2.0 km\,s$^{-1}$ ($t$ = 15 ps), and (d) 3.0 km\,s$^{-1}$ ($t$ = 13 ps). }
  \label{deform}
\end{figure*}

\section{Results and discussion}
The NEMD simulations of shock loading in nanopowdered Mg (NP-Mg) are performed for a range of particle velocities from 0.5 to 3.0 km\,s$^{-1}$. Upon shock, wave propagation and interactions undergo series of stages in NP-Mg: the impact-induced shocks, subsequent release fans originating at free surfaces, and interaction of the opposing release fans, which yield well-defined shock compression, release, tension, and spallation. Upon impact, NP-Mg undergoes the plasticity, governed via stacking faults deformation twinning at $u_{\rm p}$ $\le$ 1.0 km\,s$^{-1}$, while solid disordering or shock-melting at $u_{\rm p}$ $\ge$ 2.0 km\,s$^{-1}$. Prompting by plastic deformation, consolidation is following, i.e., the preexisting pores are completely filled, yielding a columnar nanocrystalline structure with grainboundaries (GBs) and GB tripple junctions. 

\subsection{Compression stage}
\subsubsection{Shock consolidation: ``heterogeneous'' plasticity and ``homogeneous'' disordering}
Upon impact, the shock wave along the $x$-direction prompts an apparent consolidation in NP-Mg, contributed by both plastic deformation and pore collapse. Two different structural deformation modes are included during shock compression: i) the stacking faults (SF) and deformation twinning at lower impact velocities ($u_{\rm p}$ $\le$ 1.0 km\,s$^{-1}$; Figs.~\ref{deform}(a) and (b)), conducing to the grain rotation and subgrains formation, and ii) the disordering at higher impact velocities ($u_{\rm p}$ $\ge$ 2.0 km\,s$^{-1}$; Figs.~\ref{deform}(c) and (d)). Such disordering consist of the amorphization (or solid disordering) at $u_{\rm p}$ = 2.0 km\,s$^{-1}$, and the solid-liquid transition (or shock-induced melting) at $u_{\rm p}$ = 3.0 km\,s$^{-1}$. The calculated diffusion coefficients ($D$) are 10$^{-11}$--10$^{-10}$ m$^2$\,s$^{-1}$ for the former, similar to the solid crystal ($D$ $\sim$ 10$^{-10}$ m$^2$\,s$^{-1}$); whereas $D$ $\approx$ 10$^{-8}$  m$^2$\,s$^{-1}$ for the latter, approaching the liquid state ($D$ $>$ 10$^{-9}$ m$^2$\,s$^{-1}$) \cite{he14jap}.

\begin{figure}[t]
\centering
  \includegraphics[scale=0.3]{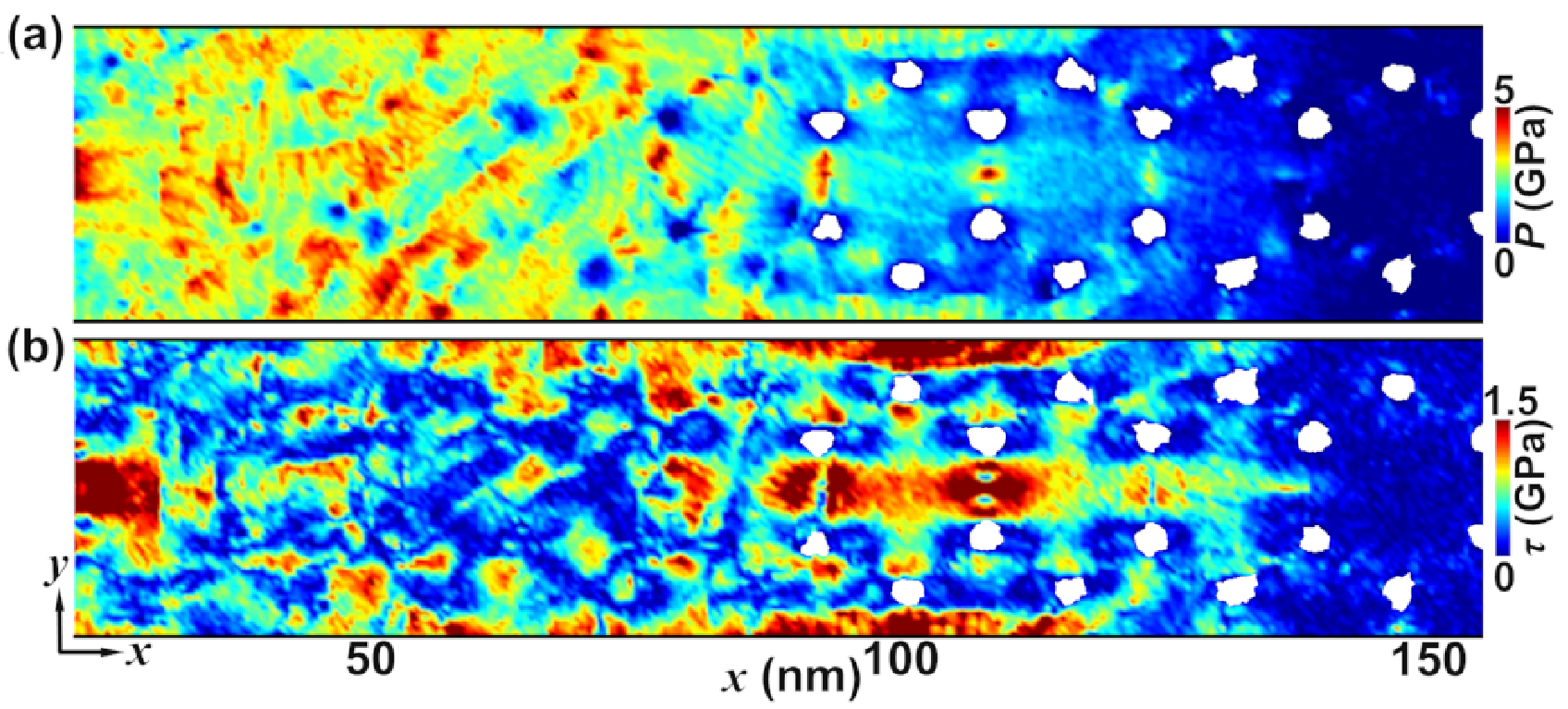}
  \caption{2D distribution maps of (a) $P(x, y)$ and (b) $\tau (x, y)$ in NP-Mg during shock consolidation at lower impact velocity ($u_{\rm p}$ = 0.5 km\,s$^{-1}$, $t$ = 23 ps).}
  \label{2dstress}
\end{figure}

The compression-induced plasticity, in facts, presents a strong dependence of the crystallographic orientation at $u_{\rm p}$ $\le$ 1.0 km\,s$^{-1}$ (Fig.~\ref{deform}(a) and (b)). The plastic deformation prefers to arise in A- and B-type grains, contributing to an apparent release of $P$ and $\tau$ locally (Fig.~\ref{2dstress}(a) and (b)), where the shock wave propagates along $x$-[1$\bar{2}$10] or the similar orientations, rather than in C- or D-type grains ([0001] or the similar orientation along $x$-axis, respectively). The ``heterogeneous'' plasticity facilitates the stress concentration or localization. Such an anisotropy of deformation becomes weak with the increase of impact velocities ($u_{\rm p}$). Upon impact, the deformation mode in NP-Mg undergoes a transition, from the ``heterogeneous'' plasticity (SFs and twins, $u_{\rm p}$ $\le$ 1.0 km\,s$^{-1}$) to ``homogeneous disordering'' ($u_{\rm p}$ $\ge$ 2.0 km\,s$^{-1}$; Fig.~\ref{deform}). 

\begin{figure}[htb]
\centering
\includegraphics[scale=0.3]{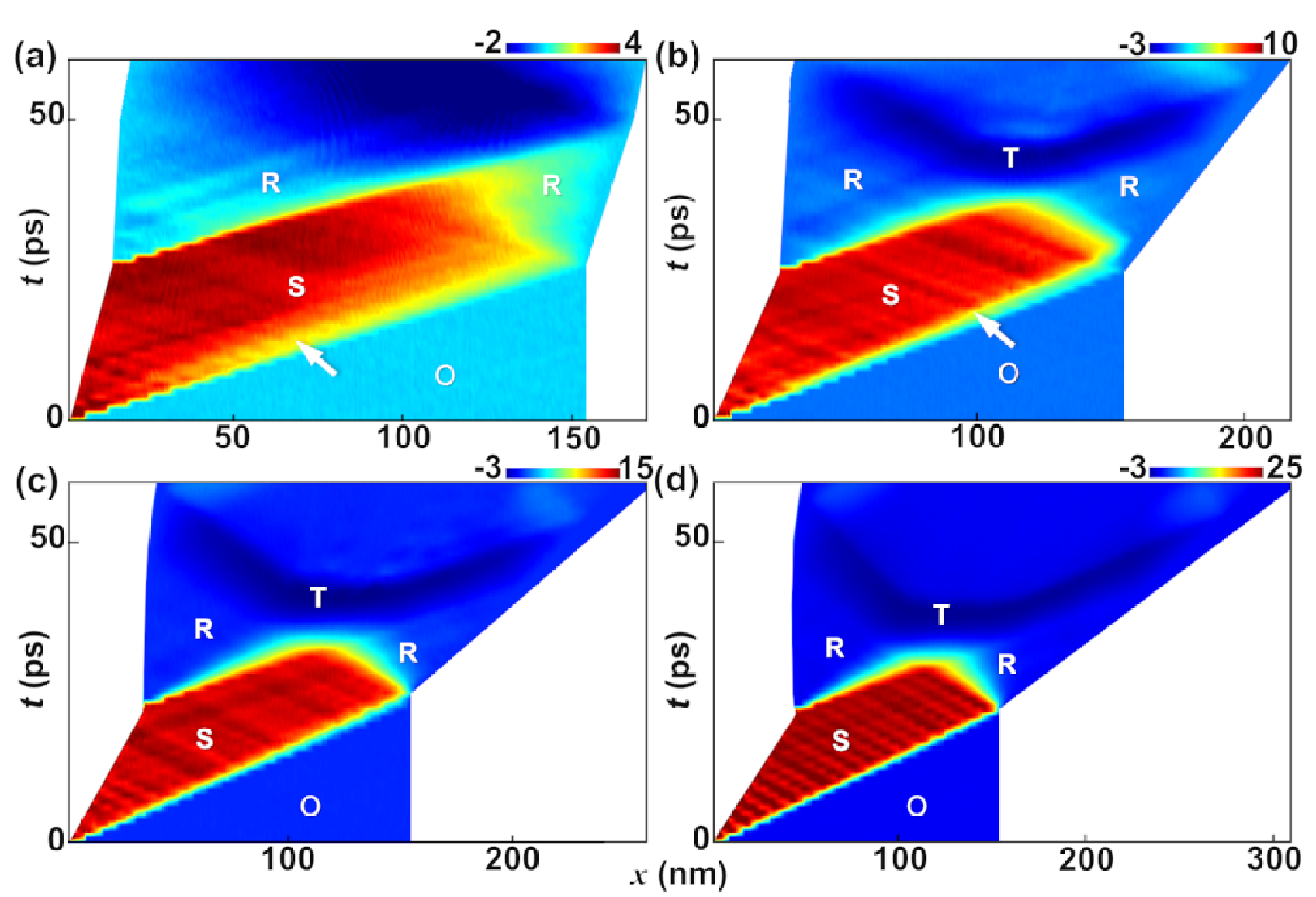}
  \caption{Position-time ($x$-$t$) diagrams showing the wave propagation and interaction in NP-Mg, at (a) $u_{\rm p}$ = 0.5 km\,s$^{-1}$, (b) 1.0 km\,s$^{-1}$, (c) 2.0 km\,s$^{-1}$, and (d) 3.0 km\,s$^{-1}$, respectively. The color coding is based on pressure $P$ (GPa). Regions O: unshocked; S: shocked; R: released; T: tension. The white arrows represent the local low-pressure regions at the plastic front.}
  \label{xt}
\end{figure}

The ``heterogeneous'' plasticity prompt an anisotropic propagation of a shock-induced plastic wave in NP-Mg, as illustrated in the traditional position-time ($x$--$t$) diagrams (Fig.~\ref{xt}). During shock compression, some low-pressure regions are observed at the plastic front at $u_{\rm p}$ $\le$ 1.0 km\,s$^{-1}$ (denoted with white arrow, Fig.~\ref{xt}(a) and (b)). These low-pressure regions are corresponding to the local release of $P$ and $\tau$ in A- and B-type grains (Fig.~\ref{deform}(a) and (b)), where the plastic deformation are activated at a low shock strength. When $u_{\rm p}$ $\ge$ 2.0 km\,s$^{-1}$, the homogeneous deformation also facilitates the uniform distribution of stress, i.e., $P$, conducing to the weakening and annihilation of low-pressure regions at the plastic shock front in the $x$--$t$ diagrams (Fig.~\ref{xt}(c) and (d)).

Contributed by the plasticity behind wave front, the nanovoids tend to be collapse at the contact points, facilitating the compaction and complete annihilation of voids. Then the nanopowder structure evolves into a deformed hexagonal columnar nanostructure at $u_{\rm p}$ $\le$ 1.0 km\,s$^{-1}$ (Fig.~\ref{deform}(a) and (b)), or a large disordering structure at $u_{\rm}$ $\ge$ 2.0 km\,s$^{-1}$ (Fig.~\ref{deform}(c) and (d)), after nanovoids compaction. Finally, these contact points extend into the continuous GBs ($u_{\rm p}$ $\le$ 1.0 km\,s$^{-1}$) or the disordering/disordering interfaces ($u_{\rm p}$ $\ge$ 2.0 km\,s$^{-1}$). Combining the primary plasticity and the subsequent nanovoid compaction, it facilitates the shock consolidation of NP-Mg, giving rise to an apparent increase of massive densities. 

Based on the simulation results above, compression-induced plasticity and nanovoid compaction accelerates the consolidation of NP-Mg. However, their underlying mechanisms are still unrevealed. Thus an explicit description of microstructure deformation, i.e., the evolutionary process and their interactions with grain boundaries and crystallographic orientation, is necessary. %Both plasticity and pore collapse, 

\subsubsection{Deformation twinning}
Deformation twinning, prompting the grain rotation partially, i.e., $\alpha$ (parent) $\to$ $\alpha^{\prime}$ (variant), is a primary deformation mode in NP-Mg during shock compression as $u_{\rm p}$ $\le$ 1.0 km\,s$^{-1}$ (Fig.~\ref{deform}(a) and (b)). Two extension twinning modes, i.e., the $\{11\bar{2}1\}\langle\bar{1}\bar{1}26\rangle$ (T$_1$) \cite{zhou19ijp, flanagan20md}, and $\{1\bar{1}02\}\langle\bar{1}101\rangle$ twins (T$_2$) \cite{wang20am, zhang20am}, and a compression twinning mode, i.e., the $\{11\bar{2}2\}\langle\bar{1}\bar{1}23\rangle$ (T$_3$) \cite{zhou19ijp} twin, as the secondary twin following the primary twin (T$_1$), are activated in NP-Mg, which also present apparent crystallographic orientation dependence. 

\begin{figure}[t]
\centering
  \includegraphics[scale=0.4]{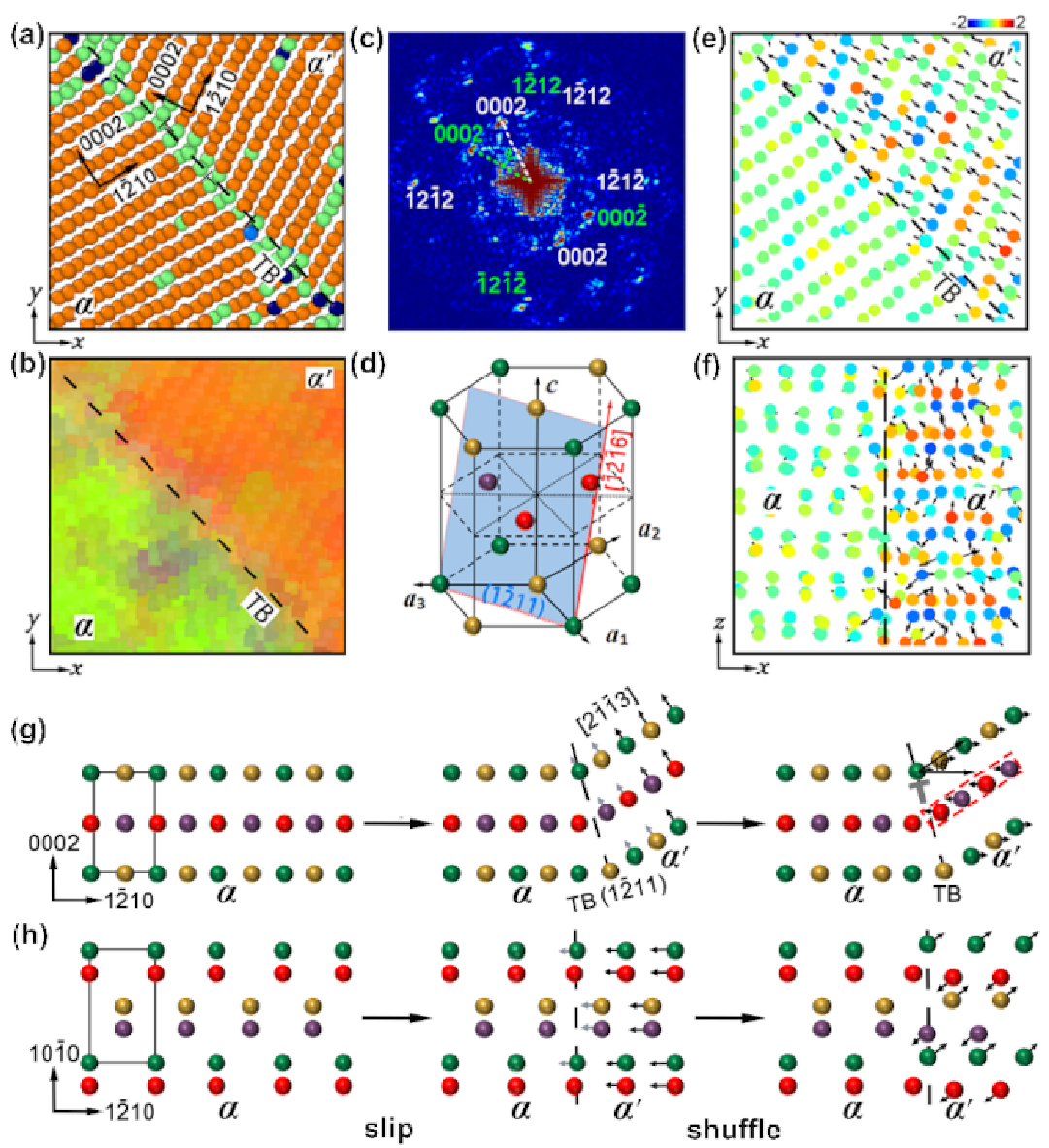}
  \caption{ (a) The atomic configuration color-coded with common neighboring analysis (CNA) and (b) the corresponding orientation map (OM) of (1$\bar{2}$11) twin. The cross sections are viewed along the $z$ axis, and 8 $\times$ 8 nm$^2$ on the $xy$ plane. (c) The corresponding diffraction patterns, where the indexes in white and green represent the parent $\alpha$ and variant $\alpha^{\prime}$, respectively. Here, the misorientation angle $\theta$ between $\alpha$ and $\alpha^{\prime}$ is about 33$^{\circ}$, similar to that in reciprocal space, i.e., the angle between dashed lines in white and green. (d) The schematic illustrations for the corresponding twinning slip systems in hcp-Mg, (e) and (f) the slip vectors (colored with $s_z$) in view of $xy$ and $xz$-planes, respectively. Arrows show the slip directions. It is noticed that the configuration at $t$ = 0 during impact is considered as the reference when calculating the slip vectors for all the cases, if no other  declarations is conducted. (g) and (h) The schematic illustrations for the nucleation mechanisms of the (1$\bar{2}1$1) twin, in terms of slip and shuffle (displacement), in view of (10$\bar{1}$0) and (0001) planes, respectively. The symbol ``$\perp$'' denotes the partial dislocation.}
  \label{t1121}
\end{figure}

{\bf {A. $\{11\bar{2}1\}\langle\bar{1}\bar{1}26\rangle$ twinning.}} When $u_{\rm p}$ $\le$ 1.0 km\,s$^{-1}$, the $\{11\bar{2}1\}\langle\bar{1}\bar{1}26\rangle$ twinning is the primary deformation mode in A- and B-type grains, as shown in Fig.~\ref{deform}(a) and (b). It contributes to a grain rotation with a rotation angle, $\theta$, (the misorientation angle between the parent and variant) of $\sim$ 33.2$^{\circ}$ around $\langle10\bar{1}0\rangle$ axis, and induces a $\alpha$ $\to$ $\alpha^{\prime}$ transition (Fig.~\ref{t1121}(a) and (b)). It is manifested via analyzing the diffraction pattern (Fig.~\ref{t1121}(c)), with the development of high temporal and spatial resolution x-ray probes. As $\{11\bar{2}1\}\langle\bar{1}\bar{1}26\rangle$ twinning proceeds, diffraction spots tend to be diffused and separated in diffraction pattern (in view of the $xy$ plane). Two characteristics of $\{11\bar{2}1\}\langle\bar{1}\bar{1}26\rangle$ twinning are presented: (i) the rotation angle between $\alpha$ and $\alpha^{\prime}$ pattern is about 32.7$^{\circ}$, consistent with the one determined in real space. The rotation angle is well agreement with the experiments results ($\theta$ $\sim$ 31--35$^{\circ}$) for $\{11\bar{2}1\}$ twins \cite{zhou19ijp, tang23jma, christian95pms}; (ii) the diffraction spots from the reciprocal space, i.e., 1$\bar{2}$11-$\alpha$ and 1$\bar{2}$1$\bar{1}$-$\alpha^{\prime}$, are coincident with each other, via analyzing the features in patterns, implying the parent ($\alpha$) and twin variant ($\alpha^{\prime}$) have the common symmetric plane $\{11\bar{2}1\}$. For $\{11\bar{2}1\}\langle\bar{1}\bar{1}26\rangle$ twin, its geometric relation can be presented in Fig.~\ref{t1121}(d), i.e., the twinning plane (11$\bar{2}$1) (dark blue) and the corresponding slip direction ($\bar{1}\bar{1}$26) (red arrow).

What is the mechanism governing such deformation twinning? The fact is that such $\{11\bar{2}1\}\langle\bar{1}\bar{1}26\rangle$ twinning is prompted by coupling the activation of $\{11\bar{2}1\}\langle\bar{1}\bar{1}26\rangle$ slip systems (Fig.~\ref{t1121}(e)) and atomic shuffles \cite{ishii16ijp} by atoms moving along the [1$\bar{1}$00] or [$\bar{1}$100] directions in the $\{0001\}$ planes (Fig.~\ref{t1121}(f)), based on the analysis of slip vectors. The atomic slips, achieved via the consecutive emission of [$\bar{1}\bar{1}$26]/6 partials on neighboring (11$\bar{2}$1) planes, facilitate the rotations from the parent hcp ($\alpha$) to the variant ($\alpha^{\prime}$) unit cell by $\sim$ 33$^{\circ}$ (Fig.~\ref{t1121}(g)), via shifting the stacking sequence from $\cdots$ABAB$\cdots$ to $\cdots$ACAC$\cdots$; the subsequent lattice reorientation via atomic shuffles in variant regions on (0001) planes (Fig.~\ref{t1121}(h)), contribute to the mirroring structure about the (11$\bar{2}$1) plane, mediating a shift from the incoherent prismatic boundary (IPB) to the coherent twin boundary (CTB: (11$\bar{2}$1) plane). 

\begin{figure}[t]
\centering
  \includegraphics[scale=0.4]{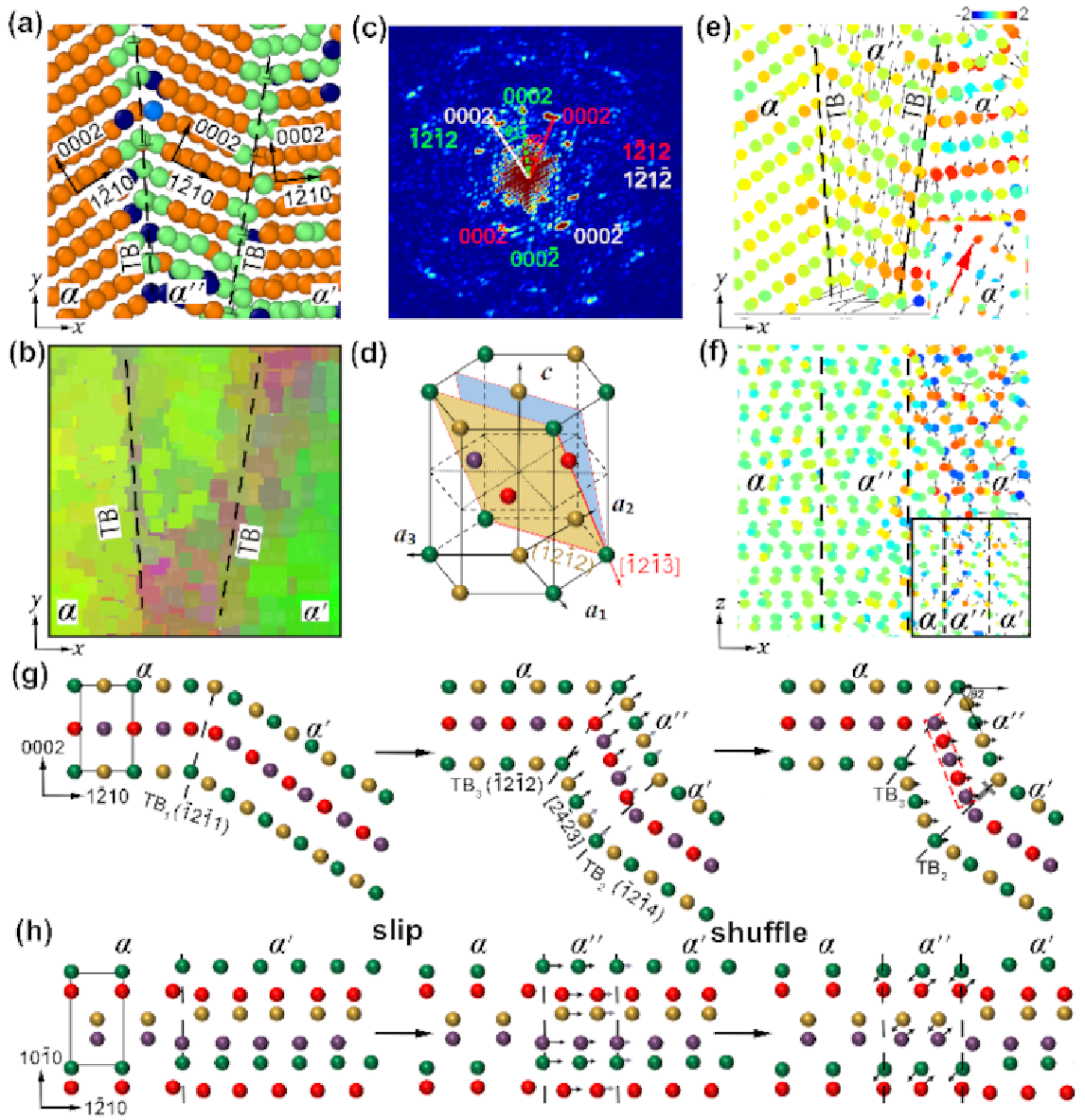}
  \caption{(a) The atomic configuration and (b) the corresponding orientation map for ($\bar{1}$2$\bar{1}$1)$\to$($\bar{1}$2$\bar{1}$2) double-twinning. The cross sections are viewed along the $z$ axis, and 5 $\times$ 5 nm$^2$ on the $xy$ plane. (c) The corresponding diffraction patterns, where the indexes in white, green and red represent the parent $\alpha$, the primary variant $\alpha^{\prime}$ (T$_1$) and the secondary variant $\alpha^{\prime\prime}$ (T$_3$). (d) The schematic illustration for the slip systems of double-twinning in hcp-Mg, (e) and (f) the slip vectors (colored with $s_z$) in view of $xy$ and $xz$-planes, respectively. The inset figure shows the slip vectors, as the configuration when only $\{11\bar{2}1\}$ twinning are activated is considered as the reference.  Arrows show the slip directions. (g) The schematic illustration for the nucleation mechanisms of the ($\bar{1}$2$\bar{1}$2) twin, in terms of slip and shuffle (displacement) in view of (10$\bar{1}$0) plane. The symbol ``$\perp$'' denotes the partial dislocation.}
  \label{t1122}
\end{figure}

\begin{figure}[t]
\centering
  \includegraphics[scale=0.4]{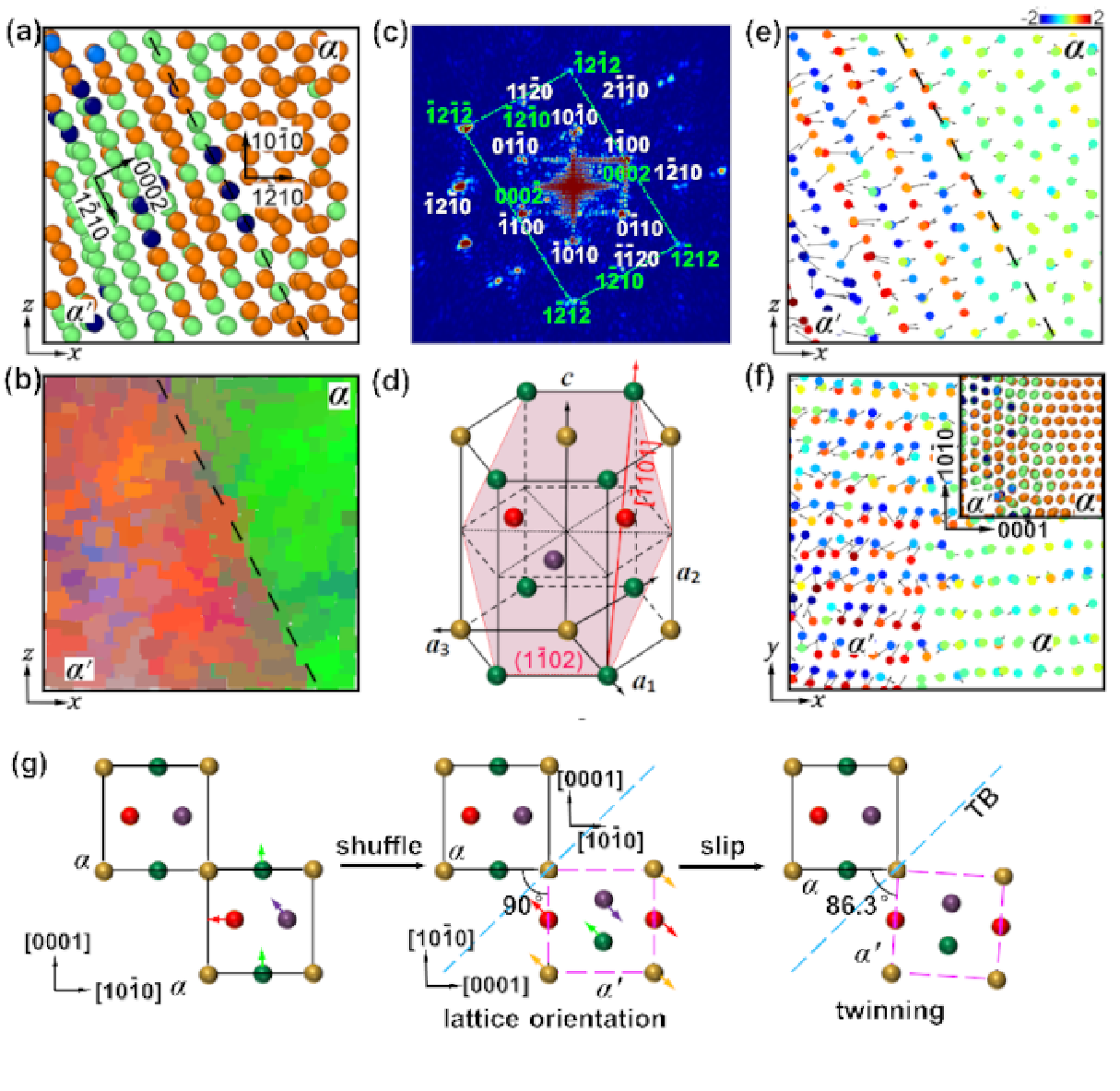}
  \caption{(a) The atomic configuration and (b) the corresponding orientation map for ($10\bar{1}$2) twinning. The cross sections are viewed along the $y$ axis, and 5 $\times$ 5 nm$^2$ on the $xz$ plane. (c) The corresponding diffraction patterns, where the indexes in white and green represent the parent $\alpha$ and the variant $\alpha^{\prime}$ (T$_2$). (d) The schematic illustration for the slip systems of ($10\bar{1}$2) twinning in hcp-Mg, (e) and (f) the slip vectors (colored with $s_z$) in view of $xy$ and $xz$-planes, respectively. Arrows show the slip directions. (g) The schematic illustration for the nucleation mechanisms of the (10$\bar{1}$2) twin, in terms of slip and shuffle (displacement) in view of (1$\bar{2}$10) plane. }
  \label{t1102}
\end{figure}

{\bf {B. $\{11\bar{2}1\}\langle\bar{1}\bar{1}26\rangle$ $\to$ $\{11\bar{2}2\}\langle\bar{1}\bar{1}23\rangle$ double twinning.}}
Following the primary $\{11\bar{2}1\}$ twins during shock impact, another compression twinning, i.e., the secondary $\{11\bar{2}2\}\langle\bar{1}\bar{1}23\rangle$ twinning, are also activated within A- and B-type grains. Then it undergoes an apparent lattice rotation, i.e., $\alpha$ (parent) $\to$ $\alpha^{\prime\prime}$ (secondary variant), adjacent to the primary variant $\alpha^{\prime}$, i.e., $\{11\bar{2}1\}\langle\bar{1}\bar{1}26\rangle$ twin  (Fig.~\ref{t1122}(a) and (b)). Along with the secondary $\{11\bar{2}2\}$ twinning, it contributes to the formation of $\{11\bar{2}1\}\langle\bar{1}\bar{1}26\rangle$ $\to$ $\{11\bar{2}2\}\langle\bar{1}\bar{1}23\rangle$ double twin. For this double-twin, its presents some distinct characteristics, also manifested in the diffraction patterns (Fig.~\ref{t1122}(c)): i) the rotation angle, between $\alpha$ and $\alpha^{\prime\prime}$, i.e., the secondary $\{11\bar{2}2\}\langle\bar{1}\bar{1}23\rangle$ twin, is about 61$^{\circ}$; ii) the misorientation angle is about 33$^{\circ}$ between $\alpha^{\prime}$ and $\alpha^{\prime\prime}$, similar to that between $\alpha$ and $\alpha^{\prime}$ when only $\{11\bar{2}1\}$ twin activates (Fig.~\ref{t1121}(c)); iii) the activation of secondary $\{11\bar{2}2\}$ twin is involved inside the primary $\{11\bar{2}1\}$ twin (Fig.~\ref{deform}(a) and (b)); and iv) all rotations induced by deformation twinning arise around the same axis, i.e., [10$\bar{1}$0]-axis ($z$-axis, Fig.~\ref{t1122}(a)--(c)). Fig.~\ref{t1122}(d) shows the geometric relation between primary twinning plane (1$\bar{2}$11) (dark blue) and secondary twinning plane (1$\bar{2}$12) (yellow), with the interaction line aligning with [10$\bar{1}$0] direction on the basal plane (0001). These results indicate that both the $\{11\bar{2}1\}$ and $\{11\bar{2}2\}$ twinning contribute to the activation of double twinning. 

What is the underlying mechanism of such double-twin, and what is the relation between the primary $\{11\bar{2}1\}$ and secondary $\{11\bar{2}2\}$ twinning? Essentially, the coupling between atomic shuffles and slips conduces to the double-twinning, similar to (11$\bar{2}$1) deformation twinning, based on the slip vector analysis (Fig.~\ref{t1122}(e) and (f)). Considering the initial configuration as the reference, similar to the postmortem analysis, it presents that the primary $\alpha^{\prime}$ is driven by $\{11\bar{2}1\}\langle\bar{1}\bar{1}26\rangle$ slips and atomic shuffles, along $\langle1\bar{1}00\rangle$ directions, while $\alpha^{\prime\prime}$ is prompted via the slips ($\{11\bar{2}2\}\langle\bar{1}\bar{1}23\rangle$ system). When the intermediate configuration (only $\alpha^{\prime}$ variant is activated within grains) is considered as the reference, $\alpha^{\prime\prime}$ is governed by atomic shuffles and slips, similar to that for $\alpha^{\prime}$ (insets, Fig.~\ref{t1122}(e) and (f)). It presents the different mechanisms for secondary twinning, i.e., $\alpha^{\prime\prime}$, based on the analysis above. Such difference can be attributed to the different analysis modes. For the former, the conclusions are obtained via directly comparing the initial and final configuration, while the dynamic process, i.e., the $\alpha$ $\to$ $\alpha^{\prime}$ precursor, which accelerates the activation of $\alpha^{\prime\prime}$ via reducing its energy barrier \cite{peng15nm}, is ignored. Then we duduce the procedures of $\{11\bar{2}1\}$ $\to$ $\{11\bar{2}2\}$ double twinning based on the discussion above (Fig.~\ref{t1122}(g)): i) prompted by atomic shuffles and slips, it first facilitates the emission and growth of $\alpha$ $\to$ $\alpha^{\prime}$, along with the primary $\{11\bar{2}1\}$ twinning; and ii) inside $\alpha^{\prime}$ variant, another $\{11\bar{2}1\}$ twinning is activated, which conduces to the rearrangement of atomic configurations, i.e., $\alpha^{\prime}$ $\to$ $\alpha^{\prime\prime}$, and the formation of $\{11\bar{2}1\}$ twin boundary (TB) between $\alpha^{\prime}$ and $\alpha^{\prime\prime}$. With the increase of secondary variant, i.e., $\alpha^{\prime\prime}$, towards the parent, it then triggers a transition of TB from the $\{11\bar{2}1\}$ to $\{11\bar{2}2\}$ between the parent ($\alpha$) and variants ($\alpha^{\prime}$ for the former, and $\alpha^{\prime\prime}$ for the latter). Simultaneously, the misorientation angle between the parent ($\alpha$) and variant changes from 33$^{\circ}$ to 61$^{\circ}$, owing to the lattice reorientation $\alpha^{\prime}$ $\to$ $\alpha^{\prime\prime}$ within the variant.

{\bf {C. $\{1\bar{1}02\}\langle\bar{1}101\rangle$ twinning}.} At the low-impact velocities ($u_{\rm p}$ $\le$ 1.0 km\,s$^{-1}$), a $\{1\bar{1}02\}\langle\bar{1}101\rangle$ twinning, is another primary plastic deformation mode in np-Mg during shock compression (Fig.~\ref{t1102}(a) and (b)), prompting a lattice rotation $\alpha$ $\to$ $\alpha^{\prime}$. Such deformation twinning can be manifested in the diffraction patterns (Fig.~\ref{t1102}(c)). For instance, it is observed that the spots for $\alpha$-parent, i.e., $\bar{1}\bar{1}$20 and 1$\bar{1}$00, are coincident with the ones for the $\alpha^{\prime}$-variant, i.e.,  $1\bar{2}$10 and 0002, respectively (in view of $xz$-plane). It presents the high similarities to the lattice reorientation with a rotation angle of 90$^{\circ}$ around the $\langle11\bar{2}0\rangle$ axis. However, there exists two distinct differences between them: (i) lattice reorientation abandons the feature of $\{1\bar{1}02\}\langle\bar{1}101\rangle$ twinning, such as the common symmetric plane $\{1\bar{1}02\}$ (Fig.~\ref{t1102}(d)); (ii) the rotation angle between $\alpha$ and $\alpha^{\prime}$ is 90$^{\circ}$ for lattice reorientation, but 86.3$^{\circ}$ for twinning. It implies such deformation twinning is strongly correlated to the 90$^{\circ}$ lattice reorientation.    

In previous studies \cite{wang19prb}, deformation twinning is considered as an intermediate state, which contributes to the lattice reorientation, as the final state, during shock impact. However, $\{1\bar{1}02\}$ twinning with misorientation angle of 86.3$^{\circ}$, can be frequently observed in HCP metals, via performing the postmorten microstructural analysis using SEM and TEM \cite{bhattacharyya09am, zhang12sm, yu14mrl, zhou19ijp}. It implies that the $\{1\bar{1}02\}$ twinning is a stable rather than transient phase during shock-induced plasticity. Furthermore, the discussion above also presents a sequence in shock-induced deformation in HCP-Mg, i.e., the preferential lattice reorientation and the following $\{11\bar{2}1\}\langle\bar{1}\bar{1}26\rangle$ twinning. Consequently, it is then deduced that $\{1\bar{1}02\}$ deformation twinning is contributed by coupling the transient lattice reorientation and slips. Upon impact, atomic shuffles, by atoms moving along $\langle10\bar{1}1\rangle$ directions in the $\{10\bar{1}2\}$ planes, based on the analysis of slip vectors (Fig.~\ref{t1102}(e) and (f)), first prompt the apparent rotations of the parent $\alpha$ by 90$^{\circ}$, around $\{1\bar{2}10\}$ direction (Fig.~\ref{t1102}(g)); and the subsequent emission of (10$\bar{1}$2)[10$\bar{1}$1] slips facilitates the lattice reorientation to deformation twinning, with a misorientation angle of 86.3$^{\circ}$ (Fig.~\ref{t1102}(g)). 

\begin{figure}[t]
\centering
  \includegraphics[scale=0.4]{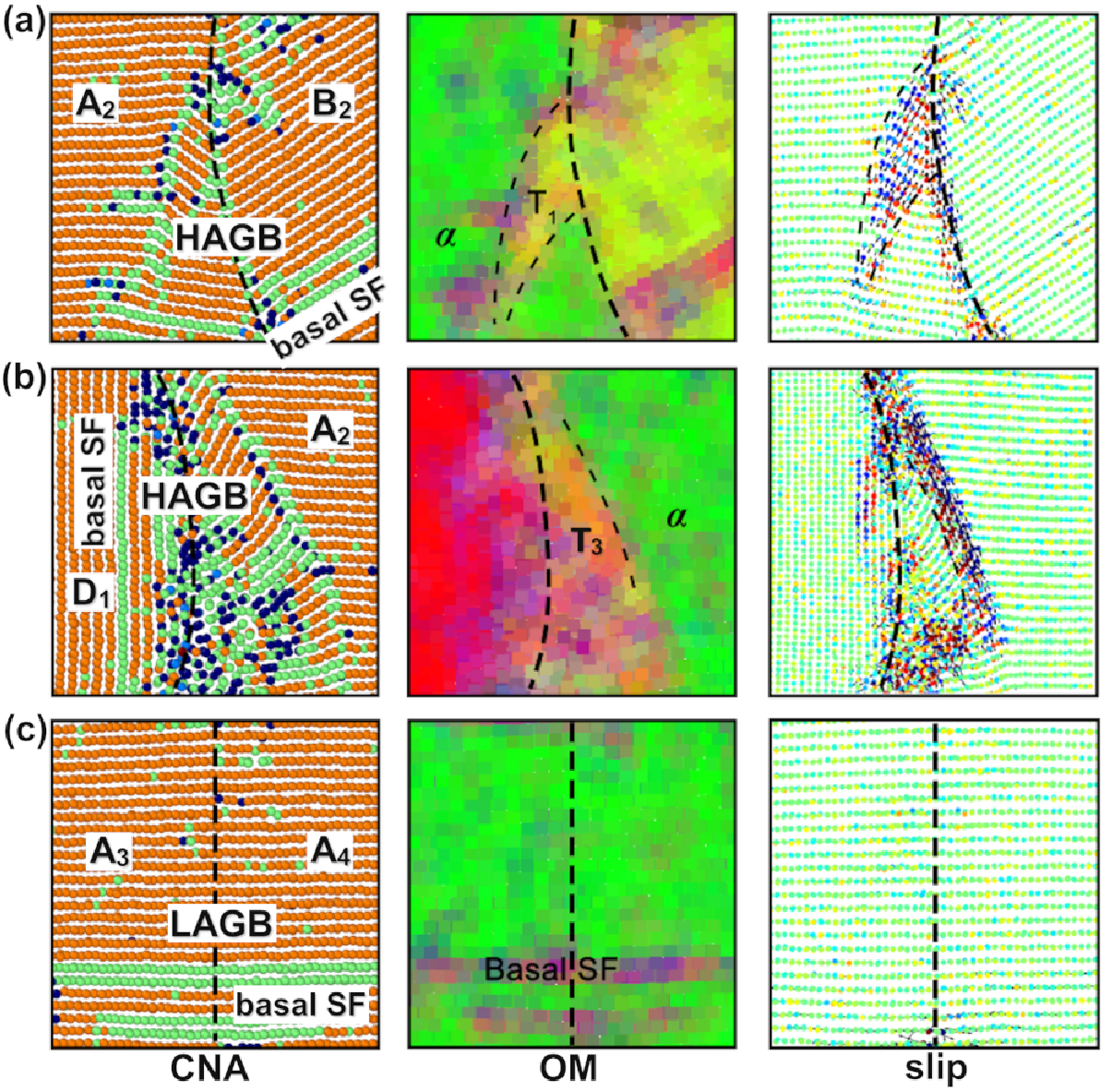}
  \caption{Microstructural configurations showing plasticity in NP-Mg during shock impact, mediated by (a) and (b) high angle grain boundaries (HAGBs) and (c) the low-angle grain boundaries (LAGBs). The color-coding are based on common neighboring analysis (CNA), orientation maps (OM), and slip vectors, $s_z$ (The colorbar is ranged from -2.0 to 2.0 \AA), respectively.}
  \label{gb_twin}
\end{figure}

\subsubsection{Interfering factors of plastic deformation}
Upon shock impact, the plasticity, particularly the deformation twinning, in facts, is strongly dependent on the microstructural characteristics, i.e., the grain boundaries (GBs) and crystallographic orientation, and the shock strength, i.e., impact velocities. 

{\bf {A. Grain boundaries.}} Prompted by shock consolidations, NP-Mg tends to evolve into the nanocrystalline structure, and induces to more GBs and GB triple junctions (Fig.~\ref{deform}(a) and (b)). For plasticity, i.e., the basal stacking faults (SFs, green) and deformation twinning, in facts, it almost originates from the GBs or GB junctions, and grows towards the interior. In the simulations, the final structure, i.e., nanocrystalline Mg consists of two types of GBs, i.e., high-angle grain boundaries (HAGBs) and low-angle grain boundaries (LAGBs). For HAGBs, the shock wave first facilitates GB sliding and then mediates apparent basal SFs within B- (Fig.~\ref{gb_twin}(a)) or D-type grains (Fig.~\ref{gb_twin}(b)), and $\{11\bar{2}1\}$ (T$_1$, Fig.~\ref{gb_twin}(a)) or $\{11\bar{2}2\}$ (T$_3$, Fig.~\ref{gb_twin}(b)) deformation twinning within A-type grains, along with the release of stresses (Fig.~\ref{2dstress}). Interestingly, the activation of $\{11\bar{2}2\}$ deformation twinning and stacking faults, accelerate the GB roughening, giving rise to a broader HAGB (Fig.~\ref{gb_twin}(b)). For LAGBs, i.e., GBs between two A-type grains, no GB sliding and deformation twinning arise owing to the stability of GBs, and only the basal SFs, originate from the GB and grow towards the interior of A-type grains, under shock loading (Fig.~\ref{gb_twin}(c)).

\begin{figure}[t]
\centering
  \includegraphics[scale=0.3]{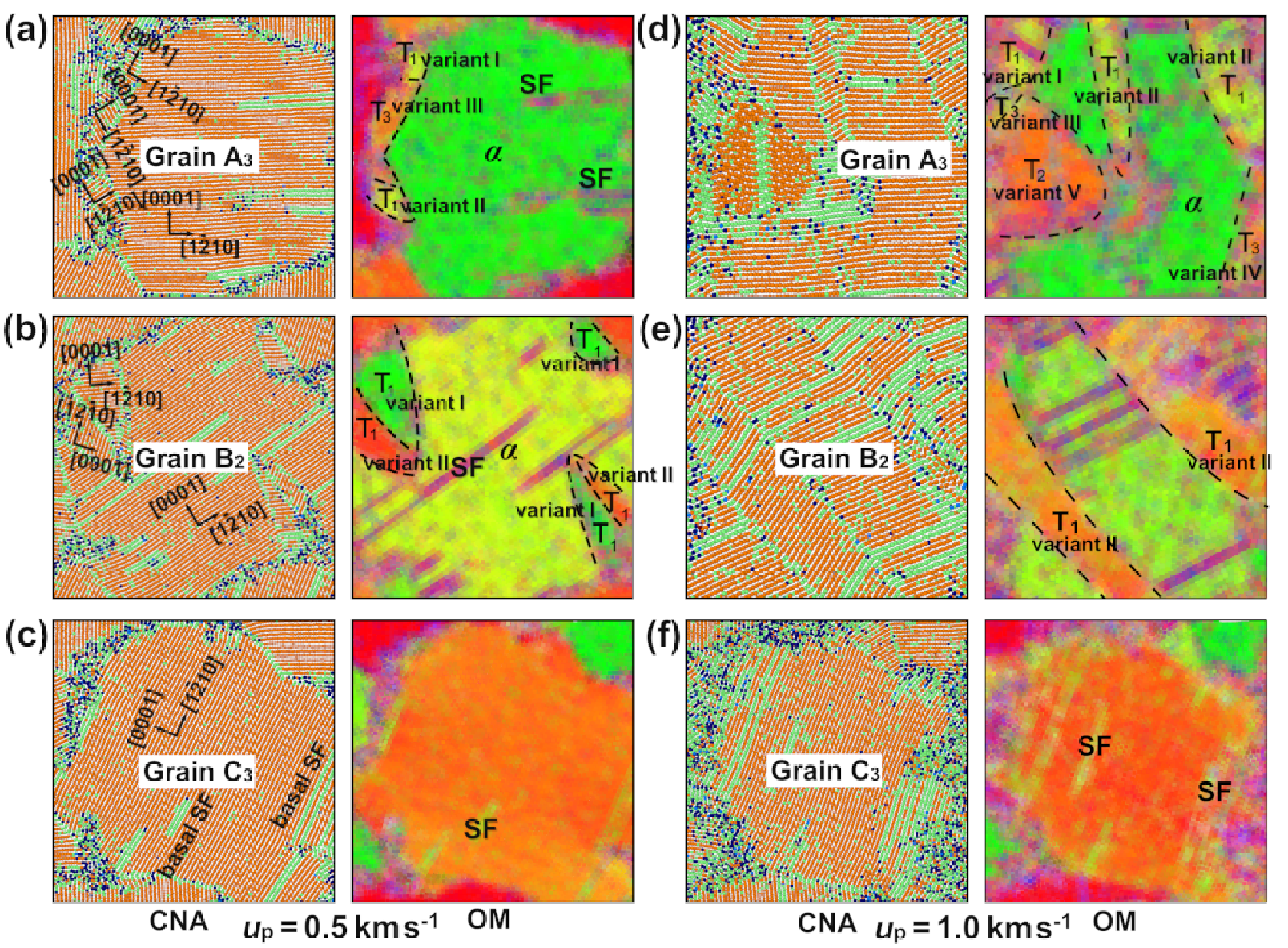}
  \caption{Microstructural configurations showing the plasticity (deformation twinning), within A-, B-, and C-type grains, in NP-Mg during shock compression, at (a)--(c) $u_{\rm p}$ = 0.5 and (d)--(f) 1.0 km\,s$^{-1}$, respectively. The color-coding are based on common neighboring analysis (CNA) and orientation maps (OM), respectively. }%The insets show the corresponding diffraction patterns, and the indexes in white and green represent the parent $\alpha$ after grain rotation and variant $\alpha^{\prime}$ induced by shock impact, respectively, and the ones in red are corresponding to the initial configuration after thermal relaxation. }
  \label{orient_twin}
\end{figure}

\begin{figure}
\centering
  \includegraphics[scale=0.25]{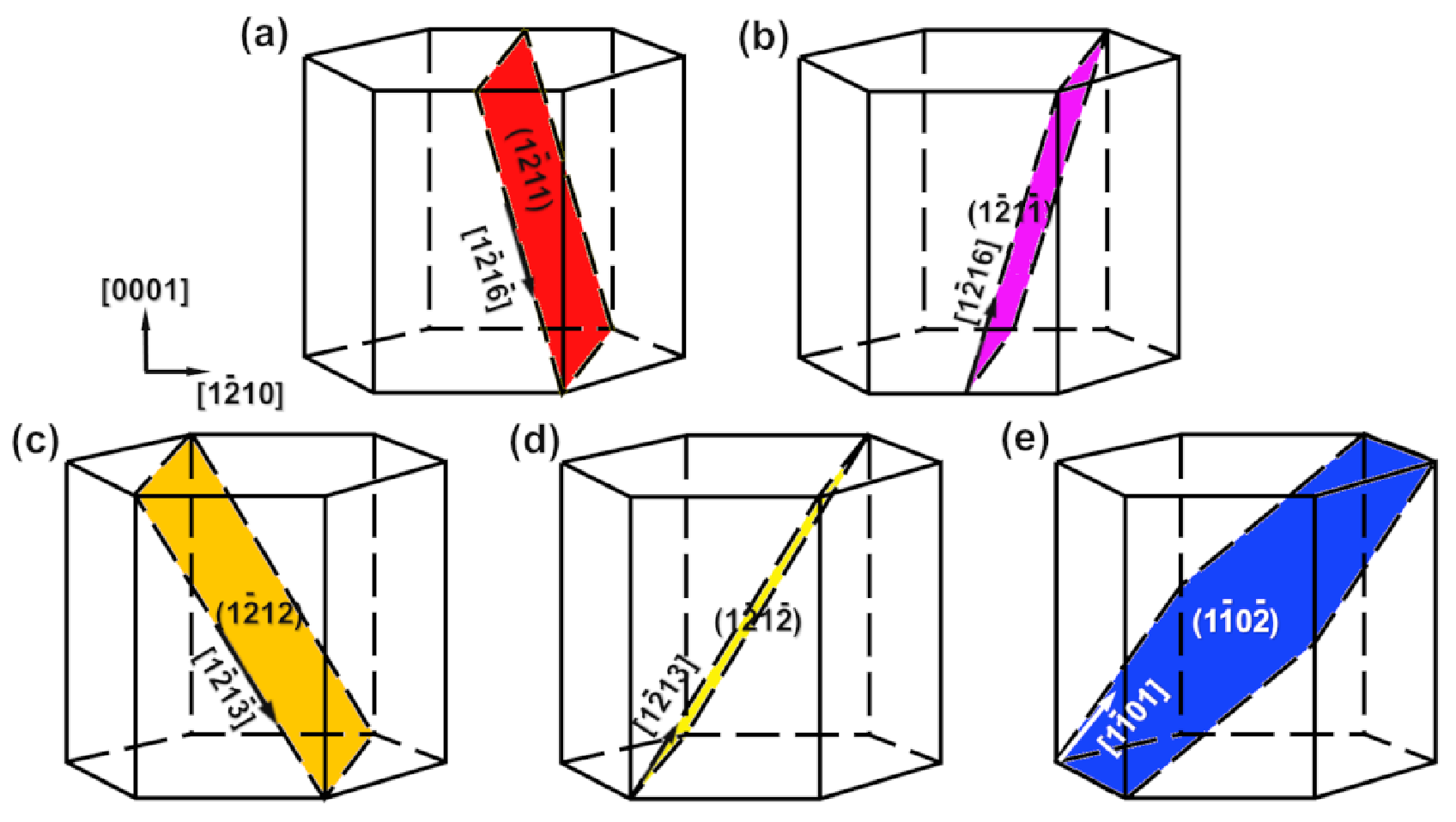}
  \caption{Schematic illustrations for five slip system for deformation twins in NP-Mg during shock compression: (a) $(1\bar{2}11)[1\bar{2}1\bar{6}]$, (b) $(1\bar{2}1\bar{1})[1\bar{2}16]$, (c) $(1\bar{2}12)[1\bar{2}1\bar{3}]$, (d) $(1\bar{2}1\bar{2})[1\bar{2}13]$, and (e) $(1\bar{1}0\bar{2})[1\bar{1}01]$, respectively.}
  \label{slip}
\end{figure}

\begin{table*}[t]
\centering
\caption{Resolved shear stress ($\tau_{\rm RSS}$: GPa) in shocked NP-Mg, for selected twinning slip systems within the grains with different crystallographic orientation at $u_{\rm p}$ = 0.5 and 1.0 km\,s$^{-1}$, respectively. Y and N represent the activated and non-activated twinning slip systems, respectively.}
\label{rss}
%\begin{threeparttable} 
%\setlength{\tabcolsep}{2mm}{
\begin{tabular}{cccccccccc}
\hline\hline
\\
        &    &  \multicolumn{2}{c}{A-type Grain} & \multicolumn{2}{c}{B-type Grain} & \multicolumn{2}{c}{C-type Grain} & \multicolumn{2}{c}{D-type Grain} \\
    $u_{\rm p}$ & Twinning slip system & $\tau_{\rm RSS}$ & Emission & $\tau_{\rm RSS}$ & Emission & $\tau_{\rm RSS}$ & Emission & $\tau_{\rm RSS}$ & Emission  \\
%           &      &             & (GPa) & \\
 \hline
 \\
\multirow{3}{*}{0.5} & (1$\bar{2}$11)[1$\bar{2}1\bar{6}$] (variant I) & 3.26 & Y & 2.86 & Y & 1.44 & N & 1.20 & N \\
    & ($1\bar{2}1\bar{1}$)[$1\bar{2}16$] (variant II) & 3.05 & Y & 2.20 & Y & 1.09 & N & 1.31 & N \\
    & (1$\bar{2}$12)[1$\bar{2}$1$\bar{3}$] (variant III) & 2.93 & Y & 1.70 & N & 1.44 & N & 0.93 & N \\\\

\multirow{5}{*}{1.0} &  (1$\bar{2}$11)[1$\bar{2}1\bar{6}$] (variant I) & 4.52 & Y & 3.47 & N & 1.94 & N & 1.72 & N \\
    & ($1\bar{2}1\bar{1}$)[$1\bar{2}16$] (variant II) & 4.30 & Y & 3.60 & Y & 1.59 & N & 1.81 & N \\
    & (1$\bar{2}$12)[1$\bar{2}$1$\bar{3}$] (variant III) & 3.57 & Y & 1.95 & N & 1.81 & N & 1.17 & N \\  
    & ($1\bar{2}1\bar{2}$)[$1\bar{2}13$] (variant IV) & 2.66 & Y & 1.66 & N & 0.93 & N & 1.07 & N \\
    & (1$\bar{1}0\bar{2}$)[1$\bar{1}$01] (variant V) & 2.50 & Y & 1.58 & N & 1.93 & N & 0.80 & N \\\\
\hline\hline
\end{tabular}
\end{table*}

{\bf {B. Crystallographic orientation.}}
The plasticity also presents an apparent crystallographic orientation dependence, based on the results above. For instance, it is observed that deformation twinning prefers to arise within A- and B-type grains, and basal SFs are predominated within C- and D-type grains, during shock compression at $u_{\rm p}$ = 0.5 km\,s$^{-1}$ (Fig.~\ref{orient_twin}(a)--(c)). For deformation twinning, it then prompts three different types of rotated [10$\bar{1}$0] $\alpha^{\prime}$ variants within A-type grains (variants I--III, Fig.~\ref{orient_twin}(a)), but two types of rotated [10$\bar{1}$0] $\alpha^{\prime}$ variant within B-type grains (variants I--II, Fig.~\ref{orient_twin}(b)). Such difference can be illustrated that the shock forces $\alpha$ to $\alpha^{\prime}$ via the activation of three twinning slip systems within A-type grains, i.e., ($1\bar{2}$11)$[1\bar{2}1\bar{6}]$ (T$_1$), (1$\bar{2}$11)[1$\bar{2}1\bar{6}$] (T$_1$), and $(1\bar{2}12)[1\bar{2}1\bar{3}]$ (T$_3$), respectively, while only the first two twinning slip systems within B-type grains. Between the variants I and II, the misorientation angle is rounded to $\sim$ 66$^{\circ}$. The boundaries between two rotated variants is the coherent twin boundary (CTB).

Such orientation effects on plasticity can be explained with resolved shear stress ($\tau_{\rm RSS}$, in Table~\ref{rss}), which is a key quantify for describing the twinning nucleation. Upon shock compression ($u_{\rm p}$ = 0.5 km\,s$^{-1}$), three different twinning slip systems are activated in np-Mg: ($1\bar{2}$11)$[1\bar{2}1\bar{6}]$ (Fig.~\ref{slip}(a)), (1$\bar{2}$11)[1$\bar{2}1\bar{6}$] (Fig.~\ref{slip}(b)), and $(1\bar{2}12)[1\bar{2}1\bar{3}]$ (Fig.~\ref{slip}(a)). Their values of $\tau_{\rm RSS}$ are different in the grains with different crystallographic orientation. Within A-type grains (shock along $[1\bar{2}10]$ direction), $\tau_{\rm RSS}$ for all three slip systems are much larger, i.e., $\tau_{\rm RSS}$ $>$ 2.0 GPa, so three different variants $\alpha^{\prime}$ (variants I--III) are observed. Within B-type grains, the value of $\tau_{\rm RSS}$ for slip system $(1\bar{2}12)[1\bar{2}1\bar{3}]$ is smaller than the other two slips system (1.70 $<$ 2.20 or 2.86 GPa), implying that the $(1\bar{2}12)[1\bar{2}1\bar{3}]$ twinning slip system is more difficult to activate in B-type grains. Within C- and D-type grains, all values of $\tau_{\rm RSS}$ for three twinning slip systems are much smaller ($\tau_{\rm RSS}$ $<$ 2.0 GPa), and thus no $\alpha$ $\to$ $\alpha^{\prime}$ arises.

{\bf {C. Impact velocities.}} Based on the discussion above, it is deduced that increasing $\tau_{\rm RSS}$ is one of most effective methods, to stimulate the activation of deformation twinning. Raising the impact strength or impact velocities ($u_{\rm p}$), it prompts an apparent increase of the corresponding $\tau_{\rm RSS}$ for each slip system. We then study the effect of impact velocities on shock-induced deformation. Within A-type grain, shock impact at $u_{\rm p}$ = 1.0 km\,s$^{-1}$ conduces to two new type of variants, apart from variants I--III activated at $u_{\rm p}$ = 0.5 km\,s$^{-1}$, i.e., the variant IV (a type of rotated [10$\bar{1}$0] $\alpha^{\prime}$ variant), owing to the emission of $(1\bar{2}1\bar{2})[1\bar{2}13]$ twinning, and the variant V (a type of rotated [0001] variant), contributed by the $(1\bar{1}0\bar{2})[1\bar{1}01]$ twinning (Fig.~\ref{orient_twin}(d)). Such phenomena can be interpreted via calculating their corresponding resolved shear stress, i.e., $\tau_{\rm RSS}$ $\ge$ 2.50 GPa (Table~\ref{rss}). Within B-type grain, the higher impact velocity ($u_{\rm p}$ = 1.0 km\,s$^{-1}$) just prompts one type of variant (variant II, Fig.~\ref{orient_twin}(e)), but two types (variant I and II) at $u_{\rm p}$ = 0.5 km\,s$^{-1}$. However, the calculated $\tau_{\rm RSS}$ cannot support this result, where $\tau_{\rm RSS}$ > 2.50 GPa for $(1\bar{2}11)[1\bar{2}1\bar{6}]$ twinning at $u_{\rm p}$ = 1.0 km\,s$^{-1}$ ($\tau_{\rm RSS}$ = 3.47 GPa, Table~\ref{rss}). An explanation is proposed that the final twinning band, owing to rapid growth of primary variant II, prevents the formation of other variant, and thus no variant I is observed. %Increasing $u_{\rm p}$ also facilitates the 

%During shock impact, it also facilitates an apparent shift of crystallographic orientation via dislocation slips and deformation twinning, which is significantly dependent on $u_{\rm p}$. 

\subsubsection{Shock-induced nanovoid compaction}
Under shock compression, the nanovoid compactions, i.e., local densification, are involved in the consolidation of NP-Mg. Such compaction prompts the structural deformation, including the plasticity at lower impact velocities ($u_{\rm p}$ $\le$ 1.0 km\,s$^{-1}$) and disordering at higher impact velocities ($u_{\rm p}$ $\ge$ 2.0 km\,s$^{-1}$). Reversely, the activities of structural deformation also conduces to the void compactions induced by the local stress-concentration. To understand the correlations between structural deformation and void compactions, an explicit description of procedures of compactions in dynamics is necessary.

\begin{figure*}[t]
\centering
  \includegraphics[scale=0.7]{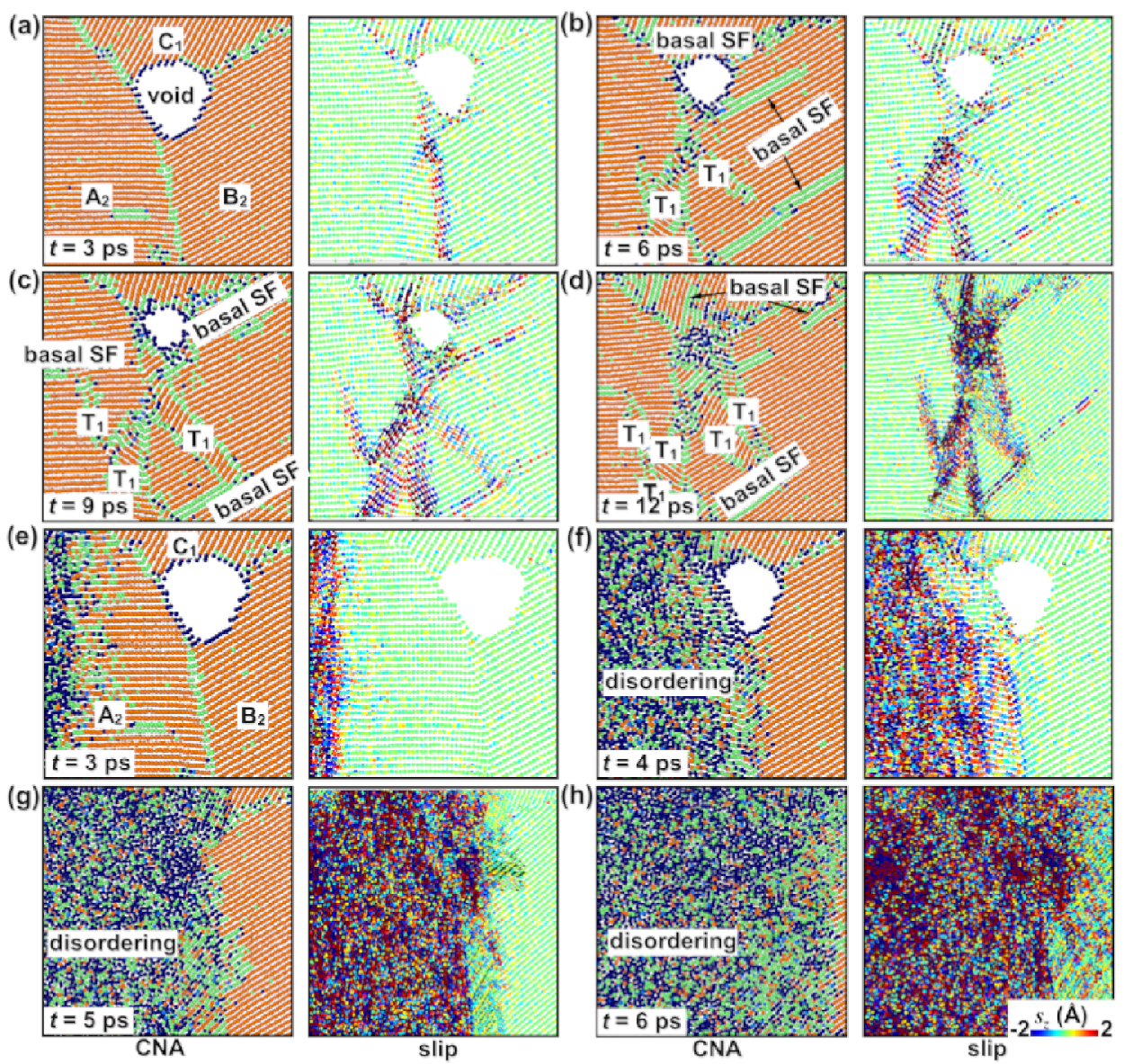}
  \caption{Snapshots of nanovoids compaction for NP-Mg during shock impact, induced by two mechanisms: (a)--(d) the plasticity, i.e., deformation twinning and basal SFs at $u_{\rm}$ = 0.5 km\,s$^{-1}$, and (e)--(h)  disordering, owing to the catastrophic activation of dislocation with different slip systems at $u_{\rm p}$ = 2.0 km\,s$^{-1}$, respectively. The color-coding is based on common neighboring analysis (CNA) and slip vectors, respectively. }
  \label{densification}
\end{figure*}

We here explore the dynamic procedures of nanovoid compactions locally, induced by plasticity (Fig.~\ref{densification}(a)--(d)) and disordering (Fig.~\ref{densification}(e)--(h)), during shock compression. At the lower-impact velocities (e.g., $u_{\rm p}$ = 0.5 km/s), GBs, adjacent to void, prefer to slide when shock wave passes through, giving rises to apparent the stress concentration ($t$ = 3 ps, Fig.~\ref{densification}(a)). Then the activation of basal SFs and deformation twinning ($\{11\bar{2}1\}$ twin) near the void, originated from GBs, conduces to the movement of some atoms within the interior of grains and GBs towards the voids ($t$ = 6 ps, Fig.~\ref{densification}(b)). By successive sliding of basal SFs and the growth of deformation twinning, more atoms are participated in the compaction of nanovoids, faciliating to the apparent void shrinkage ($t$ = 9 ps, Fig.~\ref{densification}(c)). Finally, the nanovoids are filled until they annihilate as nanovoid compaction complete ($t$ = 12 ps, Fig.~\ref{densification}(d)), contributing to a distinct increment of average mass densities. 

At higher-impact velocities, e.g., $u_{\rm p}$ = 2.0 km/s, (Fig.~\ref{densification}(e)--(h)), structural disordering, i.e., melting, due to the catastrophic, homogeneous activation of slip and their interactions, rather than the plasticity, accelerates the nanovoid compactions. Prompted by the amorphization, originated from both the GB and interior of grain ($t$ = 3 ps, Fig.~\ref{densification}(e)), it accelerates the mobility of atoms, and disordered atoms tend to be forced into the nanovoids ($t$ = 4 ps, Fig.~\ref{densification}(f)) when shock wave passes through. Then the nanovoids are compacted and finally filled by disordering atoms completely ($t$ = 5 and 6 ps, Fig.~\ref{densification}(g) and (h)). Furthering to increase impact strength, i.e, $u_{\rm p}$ = 3.0 km\,s$^{-1}$, the microjetting tends to be formed owing to the existence of nanovoids, and its growth conduces to a pronounced mass accumulations in the void, i.e., the local densification.
 
\begin{figure}[t]
\centering
\includegraphics[scale=0.4]{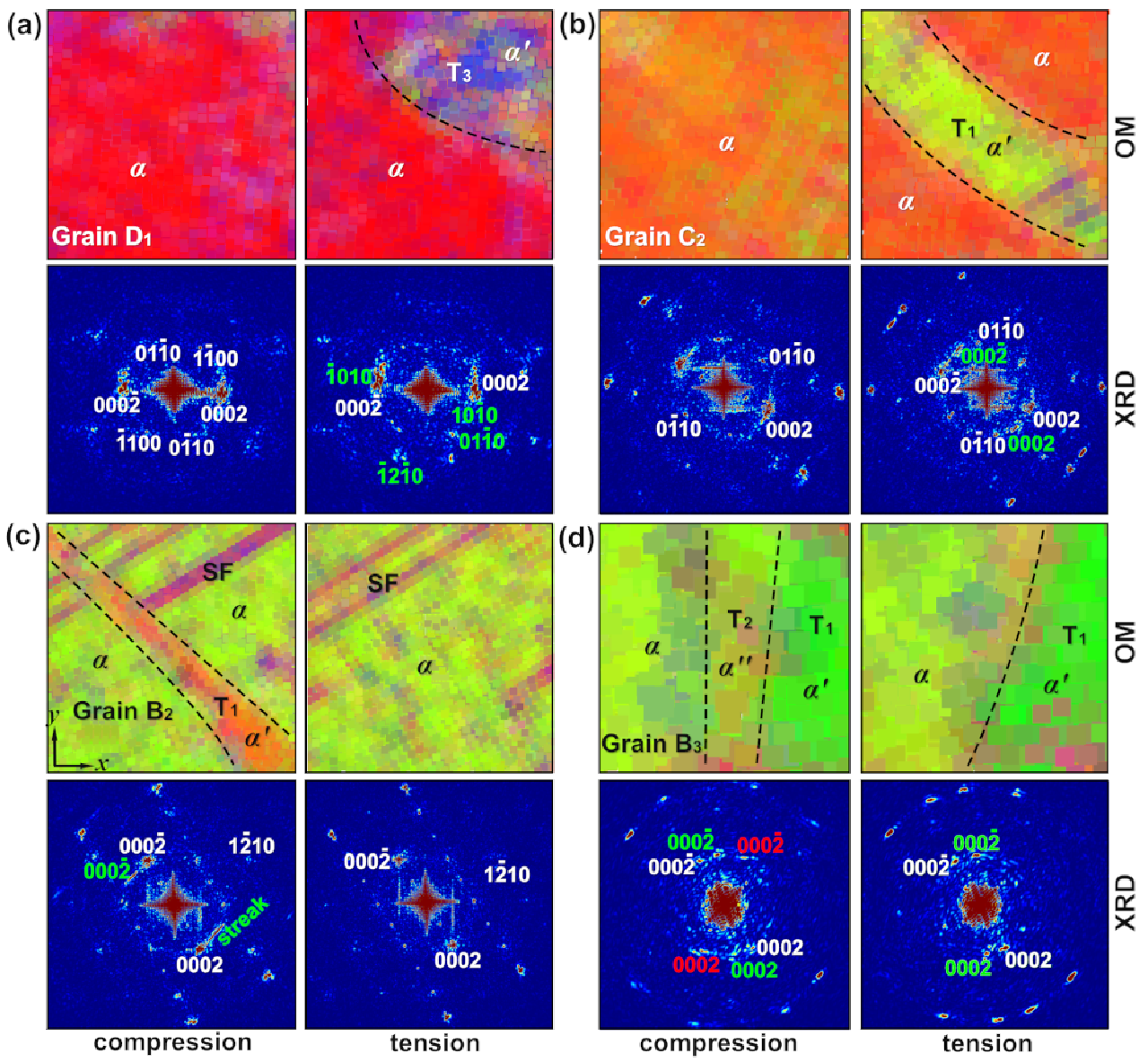}
  \caption{The orientation maps (OM) and XRD patterns, showing (a) and (b) the compression-tension asymmetry of deformation twinning within the $D$- and $C$-type grains, where deformation twinning can only be activated at tension stage; and (c) and (d) the reversible detwinning within the $B$-type grains, where deformation twinning emitted at compression stage tend to turn back at tension, for NP-Mg during shock impact ($u_{\rm p}$ = 1.0 km\,s$^{-1}$). The cross sections are 12 $\times$ 12 nm$^2$, viewed along the $z$ axis. In different patterns, the indexes, remarked in white, green, and red, represent the parents ($\alpha$), the primary twin variants ($\alpha^{\prime}$), and the secondary twin variants ($\alpha^{\prime\prime}$), respectively.}
  \label{release}
\end{figure}

\subsection{Release and tension stage}

When the shock wave travels to the free surface, it then triggers a release wave due to the shock reflection (e.g., release stage), propagating along the opposite direction (right $\to$ left), accelerating an apparent decrement of stress (R, Fig.~\ref{xt}). The subsequent interaction between two release waves, induces a tension wave within the material, causing a negative stress, i.e., the tensile stress (e.g., tension stage, T, Fig.~\ref{xt}). When the tensile stress exceeds the ultimate strength of the material, it facilitates an apparent spallation.\cite{huang12jap} However, there still exists two most significant questions in the shock dynamics: i) Is the deformation (i.e., deformation twinning) reversible when the release wave passes? What is the mechanism for the spallation caused by stretching? It is difficult to acquire the answers accurately, via analyzing the postmortem microstructures in the traditional dynamic experiments, owing to its limited information on the process.\cite{dixit17msea, kannan18jmps} %They are two of the most significant scientific inquiries in shock dynamics.

\begin{figure}[t]
\centering
\includegraphics[scale=0.3]{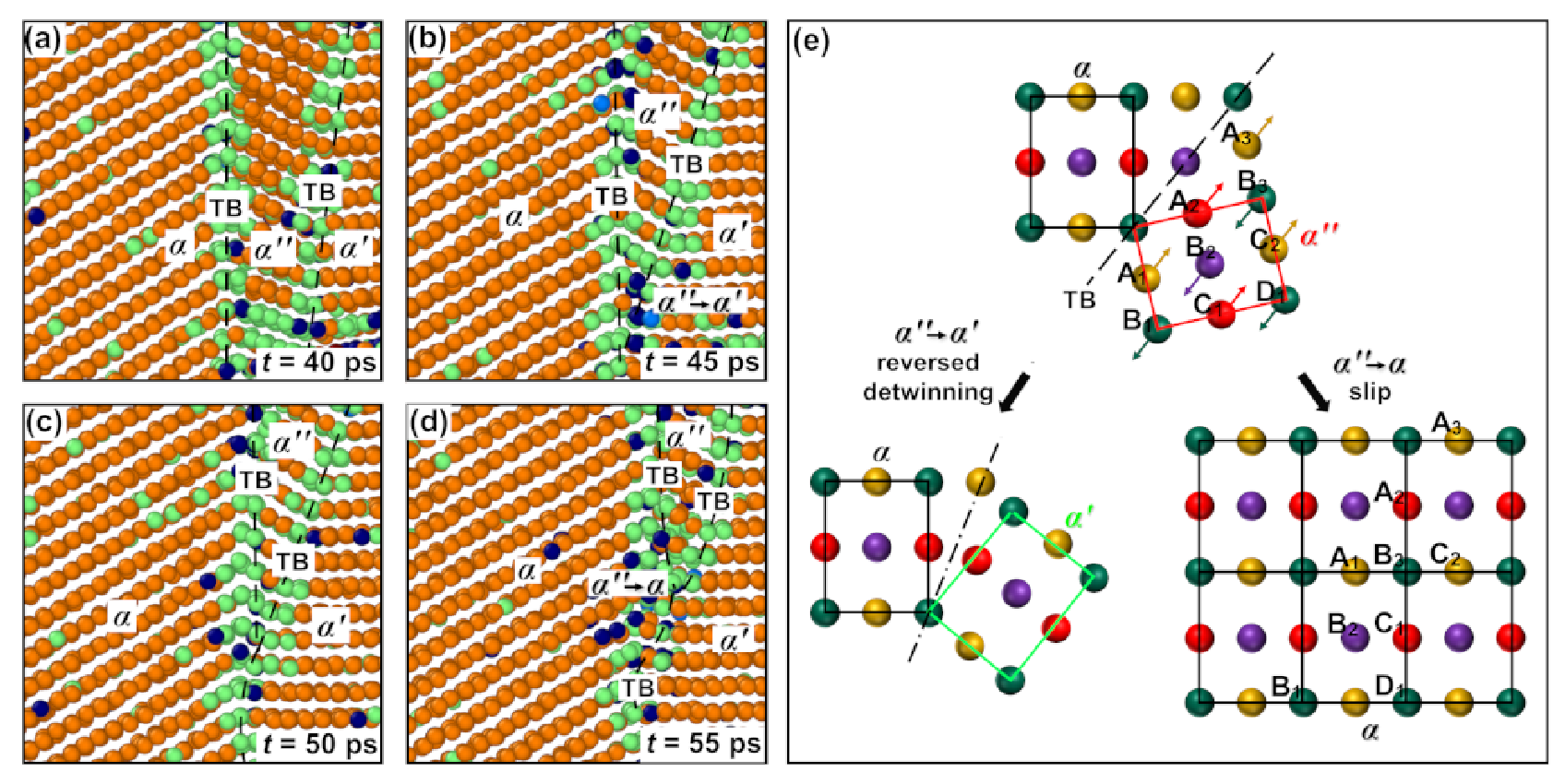}
  \caption{(a)--(d) The release-induced $\alpha^{\prime\prime}$-$\alpha^{\prime}$ reversed detwinning and the $\alpha^{\prime\prime}$-$\alpha$ transition for the $\{11\bar{2}1\}\langle\bar{1}\bar{1}26\rangle$ $\to$ $\{11\bar{2}2\}\langle\bar{1}\bar{1}23\rangle$ double twinning in NP-Mg during the release stage at $u_{\rm p}$ = 1.0 km\,s$^{-1}$. The color coding is based on common neighbor analysis (CNA). The cross sections are 12 $\times$ 12 nm$^2$, viewed along $z$-axis. Here $\alpha$, $\alpha^{\prime}$, and $\alpha^{\prime\prime}$ represent the parents, primary twin variants, and secondary twin variants, respectively. (e) The schematic illustrations for the mechanisms for $\alpha^{\prime\prime}$-$\alpha^{\prime}$ and $\alpha^{\prime\prime}$-$\alpha$ transitions.}
  \label{detwinning}
\end{figure}

\begin{figure*}[t]
\centering
  \includegraphics[scale = 0.6]{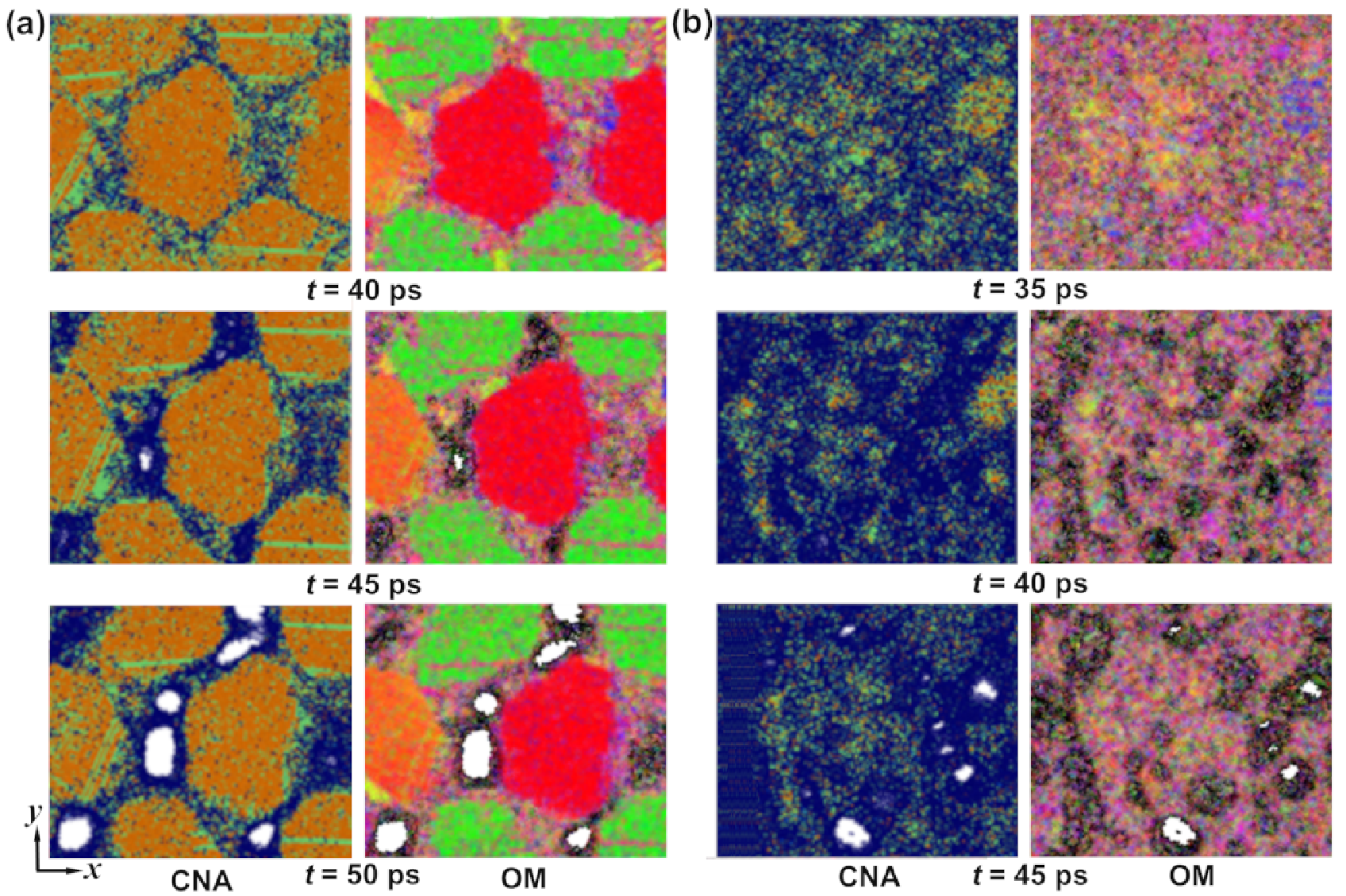}
  \caption{Deformation and spallation around the GB triple junctions during release and stages for the NP-Mg at $u_{\rm p}$ = 1.0 and 2.0 km\,s$^{-1}$, respectively. The color coding are based on common neighbor analysis (CNA) and orientation maps (OM), respectively.}
  \label{1dxrd}
\end{figure*}

\subsubsection{Twinning in tension: compression-tension asymmetry and reversible detwinning}
We here first focus on the plasticity, especially the deformation twinning, in NP-Mg during the release and tension. 

{\bf {A. Compression-tension asymmetry of deformation twinning.}} From the previous section, it is evident that certain pairs of twins arise within certain grains under shock. Is the deformation twin crystals' evolution process reversible during unloading or stretching? In the study on shock deformation of single-crystal Mg with HCP structure, Chen [21] et al. used synchrotron X-ray diffraction to observe that the evolution of twins is not identical during compression and tension processes. Additionally, no twinning information was observed inside the single-crystal Mg when shocked along the [0001] grain direction (parallel to the C-axis), whereas obvious tensile twins were formed in the tensile stage. Williams [34] et al. used time-resolved in situ synchrotron X-ray diffraction shock experiments to observe significant twinning during shock compression and de-twinning during stress relief in real time in the FG AMX602 Mg alloy, a phenomenon that typically occurs during tensile twinning of $\{10\bar{1}2\}$ with a densely arranged hexagonal material structure.

{\bf {B. Reversible detwinning.}}
As for the np-Mg, we used XRD simulation methods [Figs. 14] by comparing the structural information of the shock and release stage. We found that during the relase stage, the diffraction patterns of C and D grains, which were previously difficult to produce twins, formed a new set of diffraction spots, which were analyzed to show that the C grains formed T3  [Fig. 14(a)]and the D grains formed T1 [Fig. 14(b)] which are basically consistent with the shock experiments of single crystal Mg. In contrast, the twins formed at the shock stage of A and B grains recede during release: as shown in Fig. 14 (c) and (d), the T1 twin recedes completely and the diffraction spots of the corresponding twins disappear completely at release stage. Observing the diffractograms corresponding to the T2 [Fig. 14(d)], it is found that the diffraction spots corresponding to the T2 disappear and the twin is completely retreated,  But T2's neighboring twin, T1, has not receded and has even grown. The twinning process of $\{10\bar{1}2\}$ twins is consistent with that described by Williams et al. However, the twinning process of $\{11\bar{2}1\}$ twins and $\{11\bar{2}2\}$ twins has not been confirmed by experimental observation.

\subsubsection{The spallation}
Spallation is very difficult to occur due to its low tensile stress within the np-Mg at lower impact strength (up $\ge$ 0.5 km/s). Once the value of the tensile stress exceeds the ultimate strength of the spallation, laminar cracking occurs in the np-Mg (Fig.15), with the increase of impact velocity (up = 1.0 km/s). Spallation in np-Ti primarily occurs via the softening effect 
When up = 1.0 km/s, the spallation is attributed to the GB-sliding and GB-softening at GBs. In tension, the SFs rapidly proliferate and tangle around grain boundaries [t = 40 ps; Fig.15(a)], and subsequent interaction of SFs promotes the expansion of disordering areas [t = 45 ps; Fig.15(b)], accelerating the concentration of stresses, and leading to softening of the grain boundaries and the formation of small voids. Then, these disordering experience the relative slide, leading to the widening of the grain boundaries [t = 50 ps; Fig.15(b)], and contributing to the subsequent nucleation of voids within the grain boundaries. Such relative slide at the grain boundaries induces the nucleation and growth of numerous voids, finally triggering spallation. 
The disordering within the np-Mg promotes the occurrence of spallation [Fig.15(b)], with the increase of impact velocity (up =2 km/s). During tension, the np-Mg first becomes progressively disordered at an early stage [t = 35 ps, Fig.15(b)], which then provides the conditions for void nucleation (t = 40 ps). Along with the nucleation of lots of voids in the disorder regions (t = 45 ps), the material structure eventually fails. Such disordering is essentially the result of melting leading to the softening of the local structure of np-Mg, owing to the plastic deformation in disordered regions, where dislocation slips accelerates the mechanical strength release and mobility of materials. 

\section{Summary and Conclusions}
Using NEMD, we studied the deformation and induced by shock compression and release in np-Mg. The main conclusions are listed below:

(a) Based on our findings, at a low shock intensity of 0.5 km/s, the anisotropy of shock-induced plastic wave propagation in nanopowder Mg is heavily pronounced, resulting in the formation of a plastic double-wave structure. As the shock velocity increases, the anisotropy tends to weaken or disappear.

(b) Shock induces three types of typical twins, and we focus on the deformation mechanisms of these three types of twins.. The filling of pores accompanies the shock process, and two important mechanisms for filling the pores are plasticity-induced disordered atoms and melting-driven disordered atoms.

(c) Twins are prone to nucleation at the HAGB. Three typical types of twin crystals are formed during the shock process. The nucleation and growth of twin crystals have a significant relationship with crystal orientation. The larger the angle between the shock direction and the grain C-axis, the easier twin crystals are formed, resulting in a greater abundance of inspired twin crystals. It is noteworthy that when the shock direction and the grain C-axis are parallel, no matter the intensity of the shock, twin crystals cannot be produced. In addition, rich interactions occur between twins and twins, grain boundaries, and dislocation slips.

(d) Wave release and tensile stresses promote the receding of twins and the recrystallization of some disordered atoms. Some grains formation of unloaded twins is also observed, causing orientation effects. The deformation of twins produced under shock compression exhibits high reversibility. Disorder produced under shock compression is partially reversible.

\section*{Acknowledgments}
We acknowledge the support of Natural Science Foundation (NSF) of China (Grant Nos. 11802092 and U2230401), NSF of Hunan Province (Grant Nos. 2019JJ50221, 2019JJ40127, 2020JJ5260, and 2020JJ4375), the Funding of the Hunan Education Department Project (20A248 and 22B0225), the Double first-class construction project of Hunan Agricultural University (No. SYL2019063), and the computation platform of the National Super Computer Center in Changsha (NSCC).

\bibliographystyle{apsrev}
%\bibliography{ref}

%\end{thebibliography}

\end{document}